\documentclass[aps,prb,amsmath,twocolumn]{revtex4}
\usepackage{bm}
\usepackage{graphicx}
\usepackage{amssymb}

\begin{document}

\title{Weak localization, Aharonov-Bohm oscillations and decoherence \\
in arrays of quantum dots}
\author{Dmitri S. Golubev$^1$, Andrew G. Semenov$^2$ and
Andrei D. Zaikin$^{1,2}$} \affiliation{$^1$Institute of
Nanotechnology, Karlsruhe Institute of Technology (KIT), 76021
Karlsruhe, Germany\\
$^2$I.E. Tamm Department of Theoretical Physics, P.N. Lebedev
Physics Institute, 119991 Moscow, Russia}

\begin{abstract}
Combining scattering matrix theory with non-linear $\sigma$-model and
Keldysh technique we develop a unified theoretical approach enabling one to non-perturbatively study the effect of electron-electron
interactions on weak localization and Aharonov-Bohm oscillations
in arbitrary arrays of quantum dots. Our model embraces (i) weakly disordered conductors (ii)
strongly disordered conductors and (iii) metallic quantum dots. In
all these cases at $T \to 0$ the electron decoherence time is found to
saturate to a finite value determined by the universal formula which agrees quantitatively with numerous experimental results. Our analysis provides overwhelming evidence in favor of electron-electron interactions as a universal mechanism for zero temperature electron decoherence
in disordered conductors.
\end{abstract}

\pacs{ 73.63.Kv, 73.21.La, 73.20.Fz, 73.23.}

\maketitle


\section{Introduction}

Quantum interference of electrons is a fundamentally important
phenomenon which can strongly electron transport in disordered
conductors \cite{Berg,CS,ArSh}. Quantum coherent effects are
mostly pronounced at low temperatures in which case certain
interaction mechanisms are ``frozen out'' and, hence, do not
anymore limit the ability of electrons to interfere. However,
there exists at least one mechanism, electron-electron
interactions, which remains important down to lowest temperatures
and may destroy quantum interference of electrons down to $T=0$.
In a series of papers \cite{GZ} two of the present authors
formulated a general theoretical formalism which allows to
describe electron interference effects in the presence of disorder
and electron-electron interactions at any temperature, including
the limit $T \to 0$. This approach extends Chakravarty-Schmid
description \cite{CS} of weak localization (WL) and generalizes
Feynman-Vernon path integral influence functional technique \cite{FH}
to fermionic systems with disorder and interactions. With the aid of
this technique it turned out to be possible to quantitatively
explain low temperature saturation of WL correction to conductance
$\delta G^{WL} (T)$ commonly observed in diffusive metallic wires
\cite{Moh,Gre}. It was demonstrated \cite{GZ} that this saturation
effect is caused by electron-electron interactions.

It is worth pointing out that low temperature saturation of WL
correction and of the electron decoherence time $\tau_{\varphi}$
(extracted from $\delta G^{WL} (T)$ or by other means) has been
repeatedly observed not only in metallic wires but also in
virtually any type of disordered conductors ranging from
individual quantum dots \cite{pivin} to very strongly
disordered 3d structures and granular metals \cite{BL}. Hence,
it is plausible that in all these systems we are dealing with {\it
the same} fundamental effect of electron-electron interactions.
In order to test this conjecture it is
necessary to develop a unified theoretical description which would
cover essentially all types of disordered conductors. Although the
approach \cite{GZ} is formally an exact procedure treating
electron dynamics in the presence of disorder and interactions, in
some cases, e.g., for quantum dots and granular metals, it can be
rather difficult to directly evaluate $\delta G^{WL} (T)$ within
this technique.

One of the problems in those cases is that the description in
terms of quasiclassical electron trajectories may become
insufficient, and electron scattering on disorder should be
treated on more general footing. In addition, within the approach \cite{GZ} disorder averaging is (can be) postponed until the last stage of the calculation which is convenient in certain physical
situations. In other cases -- like ones studied below -- it might
be, in contrast, more appropriate to perform disorder averaging
already in the beginning of the whole analysis. Finally, it is
desirable to deal with the model which would embrace various types
of conductors with well defined properties both in the long and short
wavelength limits.

Below we will elaborate an alternative approach which combines
the scattering matrix and Keldysh techniques with the description of electron-electron interactions in terms of quantum Hubbard-Stratonovich
fields. Note that previously a similar type of approach was employed
in order to describe Coulomb effects in tunnel junctions, see, e.g.
\cite{SZ,Z}. Here we will describe a disordered conductor by means of an array of
(metallic) quantum dots connected via junctions (scatterers) with
an arbitrary distribution of transmissions of their conducting
channels. This model will allow to easily crossover between the
limits of granular metals and those with point-like impurities and
to treat spatially restricted and spatially extended conductors
within the same theoretical framework. Electron scattering on each
such scatterer will be treated within the most general scattering
matrix formalism \cite{MB,B} adopted to include electron-electron
interaction effects \cite{Naz,GZ01,KN,GGZ03,GZ041,GZ04,BN,GGZ05}. Averaging over
disorder will be performed within the non-linear $\sigma-$model
technique in Keldysh formulation. This method has certain
advantages over the imaginary time approach  since it allows to
treat both equilibrium and non-equilibrium problems and also
enables one to include Coulomb interaction between electrons in a
straightforward manner \cite{KA}.

In this paper we will review and extend our analysis of weak localization
effects and Aharonov-Bohm oscillations in systems composed of metallic quantum dots \cite{GZ06,GZ08,GZ07,SGZ,SZ10}. In Sec. 2 we will construct
a theory for essentially non-interacting electrons including interaction
effects only phenomenologically by introducing an effective
electron dephasing time $\tau_\varphi$ as an independent parameter.  In Sec. 3  we will develop a systematic unified analysis of the effect of electron-electron interactions on weak localization and Aharonov-Bohm oscillations in both quantum dots and extended diffusive conductors. Sec. 4 is devoted to a comparison of our results with experimental observations.

\section{Weak localization in quantum dot arrays}

\subsection{The model and basic formalism}

\begin{figure}
\centerline{\includegraphics[width=9cm]{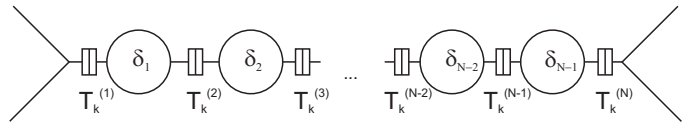}} \caption{1d
array of $N-1$ quantum dots coupled by $N$ barriers. Each quantum
dot is characterized by mean level spacing $\delta_n$. Each
barrier is characterized by a set of transmissions of its
conducting channels $T_k^{(n)}$.} \label{array1}
\end{figure}

Let us consider a 1d array of connected in series chaotic quantum dots
(Fig. \ref{array1}). Each quantum dot is characterized by its own
mean level spacing $\delta_n$. Adjacent quantum dots are connected
via barriers which can scatter electrons. Each such scatterer is
described by a set of transmissions of its conducting channels
$T_k^{(n)}$ (here $k$ labels the channels and $n$ labels the
scatterers). Below we will ignore spin-orbit scattering and
focus our attention on the case of 1d arrays. If needed, generalization of our
analysis to systems of higher dimensions can be employed in a straightforward
manner \cite{GZ06}.

An effective action $S[\check Q]$ of an array depicted in
Fig. \ref{array1} depends on
the fluctuating $4\times 4$ matrix fields \cite{GZ04,GZ06} $\check Q_n(t_1,t_2)$ defined
for each of the dots ($n=1,...,N-1$). Each of these fields
is a function of two times $t_1$ and $t_2$ and obeys
the normalization condition
\begin{equation}
\check Q_n^2=1.
\label{norm}
\end{equation}
The action of an array can be represented as a sum of two terms
\begin{eqnarray}
iS[\check Q]=iS_{d}[\check Q]+iS_{t}[\check Q].
\label{action}
\end{eqnarray}
The first term, $iS_d[\check Q]$, describes the contribution of
bulk parts of the dots. This term reads
\begin{eqnarray}
iS_d[\check Q]=\sum_{n=1}^{N-1}\frac{\pi}{\delta_n}\,{\rm Tr}\,
\left[\frac{\partial}{\partial t}\check Q_n-\alpha_n H^2\big([\check A,\check Q_n]\big)^2\right].
\label{Sd}
\end{eqnarray}
Here $H$ is an external magnetic filed,
$\alpha_n= b_n (e^2/ \hbar^2c^2)v_Fd_n^2\min\{l_e,d_n\}$,
$b_n$ is a geometry dependent numerical
prefactor \cite{B}, $d_n$ is the size
of $n-$th dot, $l_e$ is the elastic mean free path in
the dot, and $\check A$ is $4\times 4$ matrix:
\begin{eqnarray}
\check A=\left(
\begin{array}{cccc}
1 & 0 & 0 & 0 \\
0 & -1 & 0 & 0 \\
0 & 0 & 1 & 0 \\
0 & 0 & 0 & -1
\end{array}
\right).
\end{eqnarray}
The second term  in Eq. (\ref{action}), $iS_t[\check Q]$, describes electron
transfer between quantum dots. It has the form \cite{Nazarov}
\begin{eqnarray}
iS_t[\check Q]=\frac{1}{2}\sum_{n=1}^N\sum_{k}\,{\rm Tr}\,\ln
\left[1+\frac{T^{(n)}_k}{4}\big(\{\check Q_{n-1},\check Q_n\}-2\big)\right].
\label{Sj}
\end{eqnarray}
Note that here the magnetic field $H$ is included only in the term
(\ref{Sd}) describing the quantum dots while it is ignored in the
term (\ref{Sj}). Usually this approximation remains applicable at
not too low magnetic fields.

An equilibrium saddle point configuration $\check\Lambda(t_1-t_2)$
of the matrix field
$\check Q(t_1,t_2)$ depends only on the time difference and has the form
\begin{eqnarray}
\check \Lambda(t)=\int\frac{dE}{2\pi}\, e^{-iEt}\left(
\begin{array}{cccc}
-1 & 0 & 0 & 0 \\
0 & 1 & 0 & 0 \\
g^K(E) & 0 & 1 & 0 \\
0 & -g^K(E) & 0 & -1
\end{array}
\right),
\end{eqnarray}
where $g^K(E)=2[1-2f_F(E)]=2\tanh(E/2T).$
This choice of the saddle point corresponds to the following
structure of the $4\times 4$ matrix Green function $\check G$:
\begin{eqnarray}
\check G=\left(
\begin{array}{cccc}
G^A & 0 & 0 & 0 \\
0 & {\cal T}G^{A*}{\cal T} & 0 & \\
-G^K & 0 & G^R & 0 \\
0 &  {\cal T}G^{K*}{\cal T} & 0 & {\cal T}G^{R*}{\cal T}
\end{array}
\right).
\label{G}
\end{eqnarray}
Here we defined the time inversion operator ${\cal T}$:
\begin{eqnarray}
{\cal T} f(t)=f(t_f-t),
\label{T}
\end{eqnarray}
where $t_f$ will be specified later. Note that the function
$\check G$ in Eq. (\ref{G}), defined for a given disorder configuration,
should be contrasted from the Green function
\begin{eqnarray}
\check G_Q=\left[i\frac{\partial}{\partial t}+\frac{\nabla^2}{2m}
+\frac{i}{2\tau_e}\check Q\right]^{-1}
\label{GQ}
\end{eqnarray}
defined for a given realization of the matrix field $\check Q$.
In Eq. (\ref{GQ}) we also introduced the electron elastic mean free time $\tau_e$.

\subsection{Gaussian approximation}

In order to evaluate the WL correction to conductance we will
account for quadratic (Gaussian) fluctuations of the matrix field
$\check Q_n$. This approximation is always sufficient provided
the conductance of the whole sample exceeds  $e^2/h$, in certain situations
somewhat softer applicability conditions can be formulated.
Expanding in powers of such fluctuations we introduce
the following parameterization
\begin{eqnarray}
&& \check Q_n=e^{i\check W_n}\check\Lambda e^{-i\check W_n}
\nonumber\\ &&
=\check\Lambda +i[\check W_n,\check \Lambda]+\check W_n\check\Lambda\check W_n
-\frac{1}{2}\{\check W^2_n,\check\Lambda\}+{\cal O}(W^3).\;\;\;\;\;\;
\label{Q1}
\end{eqnarray}
It follows from the normalization condition (\ref{norm}) that only
$8$ out of $16$ matrix elements of $\check W$ are independent
parameters. This observation provides certain freedom to choose an
explicit form of this matrix. A convenient parameterization to
be used below is
\begin{eqnarray}
\check W_n=\left(
\begin{array}{cccc}
0 & u_{1n} & b_{1n} & 0 \\
u_{2n} & 0 & 0 & b_{2n} \\
a_{1n}+b_{1n} & 0 & 0 & v_{1n} \\
0 & a_{2n}+b_{2n} & v_{2n} & 0
\end{array}
\right).
\end{eqnarray}
With this choice the quadratic part of the action takes the form
\begin{eqnarray}
iS^{(2)}=iS^{(2)}_{ab}[a,b]+iS^{(2)}_{uv}[u,v],
\end{eqnarray}
where $iS^{(2)}_{ab}[a,b]$ does not depend on $H$
and describes diffuson modes, while $iS^{(2)}_{uv}[u,v]$ is sensitive to the
magnetic field and is responsible for the Cooperons.
The diffuson part of the action $iS^{(2)}_{ab}[a,b]$ was already analyzed before
\cite{GZ04} and will be omitted here. Below we will focus our
attention on the Cooperon contribution which reads
\begin{eqnarray}
iS^{(2)}_{uv}[u,v]&=& \sum_{n=1}^{N-1}\frac{2\pi}{\delta_n}\,{\rm Tr}\,
\bigg[ \frac{\partial}{\partial t}[u_{1n},u_{2n}]-16\alpha_nH^2\, u_1u_2\bigg]
\nonumber\\ &&
+\,\sum_{n=1}^{N-1}\frac{2\pi}{\delta_n}\,{\rm Tr}\,
\bigg[ \frac{\partial}{\partial t}[v_{2n},v_{1n}]-16\alpha_nH^2\, v_1v_2\bigg]
\nonumber\\ &&
-\,\sum_{n=1}^{N}\frac{g_n}{2}\,{\rm Tr}\,\bigg[
(u_{1n}-u_{1,n-1})(u_{2n}-u_{2,n-1})
\nonumber\\ &&
+\,(v_{1n}-v_{1,n-1})(v_{2n}-v_{2,n-1})
\bigg],
\label{quadr}
\end{eqnarray}
where $g_n=2\sum_k T_k^{(n)}=2\pi\hbar/e^2R_n$ is the dimensionless conductance
of $n-$th barrier. With the aid of the action (\ref{quadr}) we can derive
the pair correlators of the fields $u_{1,2}$ and $v_{1,2}$:
\begin{eqnarray}
\langle u_{1n}(t_1,t_2)u_{2m}(t',t'')\rangle=
\langle v_{1n}(t',t'')v_{2m}(t_1,t_2)\rangle
\nonumber\\
=\frac{\delta_m}{2\pi}\delta(t_1-t_2+t'-t'')
C_{nm}(t''-t_1),
\label{avuv}
\end{eqnarray}
where we defined a discrete version of the Cooperon $C_{nm}(t)$
obeying the equation
\begin{eqnarray}
\left(\frac{\partial}{\partial t}+\frac{1}{\tau_{Hn}}+\frac{1}{\tau_{\varphi n}}\right)C_{nm}
+\frac{\delta_n}{4\pi}\big[(g_n+g_{n+1})C_{nm}
\nonumber\\
-\,g_nC_{n-1,m}-g_{n+1}C_{n+1,m}\big]
=\delta_{nm}\delta(t).
\label{diff}
\end{eqnarray}
This equation should be supplemented by the boundary condition
$C_{nm}(t)=0$ which applies whenever one of the indices
$n$ or $m$ belongs to the lead electrode. Here $\tau_{Hn}=1/16\alpha_nH^2$ is
the electron dephasing time due to the magnetic field. In Eq. (\ref{diff})
we also introduced an additional electron decoherence time in $n-$th quantum dot $\tau_{\varphi n}$
which can remain finite in the presence of interactions. In this section we account for electron decoherence only phenomenologically by keeping the parameter
$\tau_{\varphi n}$ in the equation for the Cooperon. Rigorous description of quantum decoherence by electron-electron interactions will be carried out in Sec. 3.

\subsection{Weak localization corrections to conductance}

Let us now derive an expression for WL correction to the
conductance in terms of the fluctuating fields $u$ and $v$. In
what follows we will explicitly account for the discrete nature of
our model and specify the WL correction for a single barrier
in-between two adjacent quantum dots in the array.

We start, however, from the bulk limit, in which case
the Kubo formula for the conductivity tensor $\sigma_{\alpha\beta}$
reads
\begin{eqnarray}
&& \sigma_{\alpha\beta}(\bm{r},\bm{r}')=-i\int_{-\infty}^t dt'\; (t-t')
\nonumber\\ &&\times\,
\langle j_\beta(t',\bm{r}')j_{\alpha}(t,\bm{r})-j_{\alpha}(t,\bm{r})j_\beta(t',\bm{r}') \rangle.
\label{Kubo}
\end{eqnarray}
Following the standard procedure \cite{Berg,CS}, approximating
the Fermi function as $-\partial f_F(E)/\partial E\approx\delta(E)$ (which effectively implies taking the low temperature limit)
and using a phenomenological description of interactions as mediated by
external (classical) fluctuating fields \cite{AAK}, from Eq. (\ref{Kubo}) one can
derive the WL correction in the form:
\begin{eqnarray}
&&\delta \sigma_{\alpha\beta}^{WL}(\bm{r},\bm{r}')=-\frac{e^2}{4\pi m^2}\int_{-\infty}^t dt'\int dt''
\nonumber\\ &&\times\,
(\nabla_{\bm{r}_1}^\alpha - \nabla_{\bm{r}_2}^\alpha)_{\bm{r}_1=\bm{r}_2=\bm{r}}
(\nabla_{\bm{r}'_1}^\beta - \nabla_{\bm{r}'_2}^\beta)_{\bm{r}'_1=\bm{r}'_2=\bm{r}'}
\nonumber\\ && \times\,
\left\langle
G^R(t,\bm{r}_1; t'',\bm{r}'_2)G^A(t',\bm{r}'_1; t,\bm{r}_2)
\right\rangle_{\rm dis,\; max\; cross},\hspace{0.5cm}
\label{sigma1}
\end{eqnarray}
which implies summation over all maximally crossed
diagrams, as indicated in the subscript.
At the same time, averaging over fluctuations of $\check Q$
within Gaussian approximation is equivalent to summing
over all ladder diagrams. Since we are not going to go beyond
the above approximation, we need to convert maximally crossed diagrams
in Eq. (\ref{sigma1}) into the ladder ones. Technically this conversion
can be accomplished by an effective time reversal procedure
for the advanced Green function which can be illustrated as follows.

Consider, e. g., the second order correction to $G^A$ in the disorder potential $U_{\rm dis}(\bm{x})$
\begin{eqnarray}
&&\delta^{(2)} G^A(t',\bm{r}'_1; t,\bm{r}_2)= -i\int_{t'}^t d\tau_2\int_{t'}^{\tau_2}d\tau_1
\int d^3\bm{x_2}d^3\bm{x_1}
\nonumber\\ &&\times\,
 G^A(t',\bm{r}'_1; \tau_1,\bm{x}_1)U_{\rm dis}(\bm{x}_1)G^A(\tau_1,\bm{x}_1; \tau_2,\bm{x}_2)
\nonumber\\ &&\times\,
U_{\rm dis}(\bm{x}_2)G^A(\tau_2,\bm{x}_2; t,\bm{r}_2).
\end{eqnarray}
Making use of the property $G^A(X_1,X_2)=G^{R*}(X_2,X_1)$, we get
\begin{eqnarray}
&&\delta^{(2)} G^A(t',\bm{r}'_1; t,\bm{r}_2)= -i\int_{t'}^t d\tau_2\int_{t'}^{\tau_2}d\tau_1
\int d^3\bm{x_2}d^3\bm{x_1}
\nonumber\\ &&\times\,
 G^{R*}(t,\bm{r}_2; \tau_2,\bm{x}_2)U_{\rm dis}(\bm{x}_2)G^{R*}(\tau_2,\bm{x}_2; \tau_1,\bm{x}_1)
\nonumber\\ &&\times\,
U_{\rm dis}(\bm{x}_1)G^{R*}(\tau_1,\bm{x}_1; t',\bm{r}'_1).
\end{eqnarray}
Setting $t_f=t+t'$,
we rewrite this expression as follows
\begin{eqnarray}
&&\delta^{(2)} G^A(t',\bm{r}'_1; t,\bm{r}_2)= -i\int_{t_f-t}^{t_f-t'} d\tau_2\int_{t_f-t}^{\tau_2}d\tau_1
\nonumber\\ &&\times\,
\int d^3\bm{x_2}d^3\bm{x_1}\;G^{R*}(t_f-t',\bm{r}_2; \tau_2,\bm{x}_2)
\nonumber\\ &&\times\,
 U_{\rm dis}(\bm{x}_2)G^{R*}(\tau_2,\bm{x}_2; \tau_1,\bm{x}_1)
\nonumber\\ &&\times\,
U_{\rm dis}(\bm{x}_1)G^{R*}(\tau_1,\bm{x}_1; t_f-t,\bm{r}'_1).
\label{20}
\end{eqnarray}
Close inspection of the right hand side of Eq. (\ref{20}) allows
to establish the following relation
\begin{eqnarray}
\delta^{(2)} G^A(t',\bm{r}'_1; t,\bm{r}_2)={\cal T}\delta^{(2)}G^{R*}(t',\bm{r}_2; t,\bm{r}'_1){\cal T},
\end{eqnarray}
which turns out to hold in all orders of the perturbation theory in $U_{\rm
  dis}$. As before, the time inversion operator ${\cal T}$ is defined in Eq. (\ref{T})
with $t_f=t+t'$.

As a result, the expression for $\delta \sigma_{\alpha\beta}^{WL}$ takes the form:
\begin{eqnarray}
&& \delta \sigma_{\alpha\beta}^{WL}(\bm{r},\bm{r}')=-\frac{e^2}{4\pi m^2}\int_{-\infty}^t dt'\int dt''
\nonumber\\ &&\times\,
(\nabla_{\bm{r}_1}^\alpha - \nabla_{\bm{r}_2}^\alpha)_{\bm{r}_1=\bm{r}_2=\bm{r}}
(\nabla_{\bm{r}'_1}^\beta - \nabla_{\bm{r}'_2}^\beta)_{\bm{r}'_1=\bm{r}'_2=\bm{r}'}
\nonumber\\ && \times\,
\left\langle
G^R(t,\bm{r}_1; t'',\bm{r}'_2){\cal T}G^{R*}(t',\bm{r}_2; t,\bm{r}'_1){\cal T}
\right\rangle_{\rm dis,\; ladder}\;\;\;\;\;
\label{sigma2}
\end{eqnarray}
Rewriting Eq. (\ref{sigma2})
in terms of the matrix elements of the Green function (\ref{G}), we obtain
\begin{eqnarray}
&& \delta \sigma_{\alpha\beta}^{WL}(\bm{r},\bm{r}')=-\frac{e^2}{4\pi m^2}\int_{-\infty}^t dt'\int dt''
\nonumber\\ &&\times\,
(\nabla_{\bm{r}_1}^\alpha - \nabla_{\bm{r}_2}^\alpha)_{\bm{r}_1=\bm{r}_2=\bm{r}}
(\nabla_{\bm{r}'_1}^\beta - \nabla_{\bm{r}'_2}^\beta)_{\bm{r}'_1=\bm{r}'_2=\bm{r}'}
\nonumber\\ && \times\,
\left\langle
G_{33}(t,\bm{r}_1; t'',\bm{r}'_2)G_{44}(t',\bm{r}_2; t,\bm{r}'_1)
\right\rangle_{\rm dis,\; ladder}
\label{sigma3}
\end{eqnarray}

Our next step amounts to expressing WL correction via the Green
function $\check G_Q$ (\ref{GQ}). For that purpose
we will use the following rule of averaging
\begin{eqnarray}
&& \left\langle
G_{33}(t,\bm{r}_1; t'',\bm{r}'_2)G_{44}(t',\bm{r}_2; t,\bm{r}'_1)
\right\rangle_{\rm dis}
\nonumber\\ &&
=\left\langle
G_{33;Q}(t,\bm{r}_1; t'',\bm{r}'_2)G_{44;Q}(t',\bm{r}_2; t,\bm{r}'_1)
\right\rangle_{Q}
\nonumber\\ &&
-\, \left\langle
G_{34;Q}(t,\bm{r}_1;t,\bm{r}'_1 )G_{43;Q}(t',\bm{r}_2;  t'',\bm{r}'_2)
\right\rangle_{Q}.
\label{av}
\end{eqnarray}
One can check that within our Gaussian approximation in $u$ and
$v$ the first term in the right hand side of Eq. (\ref{av}) does
not give any contribution. Hence, we find
\begin{eqnarray}
&& \delta \sigma_{\alpha\beta}^{WL}(\bm{r},\bm{r}')=\frac{e^2}{4\pi m^2}\int_{-\infty}^t dt'\int dt''
\nonumber\\ &&\times\,
(\nabla_{\bm{r}_1}^\alpha - \nabla_{\bm{r}_2}^\alpha)_{\bm{r}_1=\bm{r}_2=\bm{r}}
(\nabla_{\bm{r}'_1}^\beta - \nabla_{\bm{r}'_2}^\beta)_{\bm{r}'_1=\bm{r}'_2=\bm{r}'}
\nonumber\\ && \times\,
\left\langle
G_{34; Q}(t,\bm{r}_1; t,\bm{r}'_1)G_{43; Q}(t',\bm{r}_2; t'',\bm{r}'_2)
\right\rangle_{Q}.
\label{sigma4}
\end{eqnarray}

Let us now turn to our model of Fig. 1 in which case the voltage
drops occur only across barriers. In this case Eq. (\ref{sigma4}),
which only applies to bulk metals, should be generalized
accordingly. Consider the conductance of an individual barrier
determined by the following Kubo formula
\begin{eqnarray}
G&=&-i\int_{-\infty}^t dt' (t-t')
\langle I(t',x')I(t,x)
\nonumber\\ &&
-\,I(t,x)I(t',x') \rangle.
\label{KuboG}
\end{eqnarray}
Here $I(t,x)$ is the operator of the total current flowing in the lead (or dot)
and $x$ is a longitudinal coordinate chosen to be in a close vicinity of the barrier. Due to the current conservation the
conductance $G$ should not explicitly depend on $x$ and $x'$.
Comparing
Eqs. (\ref{KuboG}) and (\ref{Kubo}), and making use of Eq. (\ref{sigma4})
and the relation $I(t,x)=\int d^2{\bm z}\, j_x(t,x,\bm{z}),$
where $j_x$ is the current density in the $x-$direction and $\bm{z}$ is the vector in the transversal
direction, we conclude that WL correction to the conductance of a
barrier between the left and right dots should read
\begin{eqnarray}
&& \delta G^{WL}_{LR}=\frac{e^2}{4\pi m^2}\int_{-\infty}^t dt'\int dt''\int d^2\bm{z}d^2\bm{z}'
\nonumber\\ &&\times\,
(\nabla_{x_1} - \nabla_{x_2})_{x_1=x_2=x}
(\nabla_{x'_1} - \nabla_{x'_2})_{x'_1=x'_2=x'}
\nonumber\\ && \times\,
\left\langle
G_{34; Q}(t,x_1,\bm{z}; t,x'_1,\bm{z}')G_{43; Q}(t',x_2,\bm{z}; t'',x'_2,\bm{z'})
\right\rangle_{Q}.
\nonumber\\
\label{GWL1}
\end{eqnarray}

In what follows we will assume that both coordinates
$x$ and $x'$ are on the left side from and very close to the
corresponding barrier. Let us express the Green function
in the vicinity of the barrier in the form
\begin{eqnarray}
&& \check G_Q(t,x,\bm{z};t',x',\bm{z}')=\sum_{nm}\big\{
e^{ip_nx_1-ip_mx'}\check {\cal G}^{++}_{mn}(t,t',x,x')
\nonumber\\ &&
+\,e^{-ip_nx+ip_mx'}\check {\cal G}^{--}_{mn}(t,t',x,x')
\nonumber\\ &&
+\,e^{ip_nx+ip_mx'}\check {\cal G}^{+-}_{mn}(t,t',x,x')
\nonumber\\ &&
+\,e^{-ip_nx-ip_mx'}\check {\cal G}^{-+}_{mn}(t,t',x,x')\big\}
\Phi_n(\bm{z})\Phi_m^*(\bm{z}'),
\end{eqnarray}
where $\Phi_n(\bm{z})$ are the transverse quantization modes
which define  conducting channels, $p_n$ is projection of the Fermi
momentum perpendicular to the surface of the barrier, and
the semiclassical Green function ${\cal G}_{mn}^{\alpha\beta}$ slowly varies in space.
 Eq. (\ref{GWL1}) then becomes
\begin{eqnarray}
&& \delta G^{WL}_{LR}=\frac{e^2}{4\pi m^2}\int_{-\infty}^t dt'\int dt''
\nonumber\\ &&\times\,
\sum_{mnkl}\sum_{\alpha\beta\gamma\delta=\pm 1}
(\alpha p_n-\gamma p_k)(\beta p_m-\delta p_l)
\nonumber\\ && \times\,
\left\langle
{\cal G}_{mn;34}^{\alpha\beta}(t,t,x,x'){\cal G}_{kl;43}^{\gamma\delta}(t',t'',x,x')
\right\rangle_{Q}
\nonumber\\ &&\times\,
\left. e^{i\alpha p_nx_1-i\beta p_mx'_1+i\gamma p_kx_2- i\delta p_lx'_2}
\right|_{x_1=x_2=x;x'_1=x'_2=x'}.\hspace{0.75cm}
\label{GWL2}
\end{eqnarray}

Next we require $\delta G^{WL}_{LR}$ to be independent on $x$ and $x'$, i.e.
in Eq. (\ref{GWL2}) we omit those terms, which contain
quickly oscillating functions of these coordinates.
This requirement implies that $\alpha p_n+\gamma p_k=0$
and $\beta p_m+\delta p_l=0$. These constraints in turn yield $\gamma=-\alpha$, $\delta=-\beta$,
$k=n$ and $l=m$. Thus, we get
\begin{eqnarray}
\delta G^{WL}_{LR}=\frac{e^2}{\pi m^2}\sum_{mn}\sum_{\alpha\beta=\pm 1}\int_{-\infty}^t dt'\int dt''
\alpha\beta p_n p_m
\nonumber\\  \times\,
\left\langle
{\cal G}_{mn;34}^{\alpha\beta}(t,t,x,x'){\cal G}_{nm;43}^{-\alpha,-\beta}(t',t'',x,x')
\right\rangle_{Q}.
\label{GWL3}
\end{eqnarray}

Let us choose the basis in which transmission and reflection
matrices $\hat t$ and $\hat r$ are diagonal. In this basis the
semiclassical Green function is diagonal as well, ${\cal
G}_{mn}\propto {\cal G}_{nn}\delta_{nm}$, and Eq. (\ref{GWL3})
takes the  form
\begin{eqnarray}
\delta G^{WL}_{LR}&=&\frac{e^2}{\pi}\sum_{n}\frac{p^2_n}{m^2}\int_{-\infty}^t dt'\int dt''
\nonumber\\ && \times\,
\big\langle
{\cal G}_{L,nn;34}^{++}(t,t){\cal G}_{L,nn;43}^{--}(t',t'')
\nonumber\\ &&
+\,{\cal G}_{L,nn;34}^{--}(t,t){\cal G}_{L,nn;43}^{++}(t',t'')
\nonumber\\ &&
-\,{\cal G}_{L,nn;34}^{+-}(t,t){\cal G}_{L,nn;43}^{-+}(t',t'')
\nonumber\\ &&
-\,{\cal G}_{L,nn;34}^{-+}(t,t){\cal G}_{L,nn;43}^{+-}(t',t'')
\big\rangle_{Q}.
\label{GWL4}
\end{eqnarray}
What remains  is to express the WL correction  in terms of the field
$\check Q$ only. This goal is achieved with the aid of the following
general relation \cite{GZ06}
\begin{eqnarray}
&& \delta G^{WL}_{LR}=-\frac{e^2}{\pi }\sum_{n}\int_{-\infty}^t dt'\int dt''
\nonumber\\ &&  \times\,
\big\langle
T_n \big[ v_{1L}(t,t)v_{2R}(t',t'')+v_{1R}(t,t)v_{2L}(t',t'') \big]
\nonumber\\ &&
+\, T_n^2 [v_{1L}(t,t)-v_{1R}(t,t)][v_{2L}(t',t'')-v_{2R}(t',t'')]
\big\rangle.\hspace{0.65cm}
\label{GWL5}
\end{eqnarray}
Note that the contribution linear in $T_n$, which
contains the product of the fluctuating
fields on two different sides of the barrier, vanishes identically
provided fluctuations on one side tend to zero, e.g. if the barrier is
directly attached to a large metallic lead. In contrast, the contribution
$\propto T_n^2$ in Eq. (\ref{GWL5}) survives even in this case.

Finally, applying the contraction rule (\ref{avuv}) we get
\begin{eqnarray}
 \delta G^{WL}_{LR}&=&-\frac{e^2 g}{4\pi^2 }\int_{0}^\infty dt
\big\{ \beta\big[\delta_R C_{LR}(t)+\delta_L C_{RL}(t)\big]
\nonumber\\ &&
+\,(1-\beta)\big[\delta_R C_{RR}(t)+\delta_LC_{LL}(t)\big]\big\}.
\label{GWL}
\end{eqnarray}
Here $\delta_{L,R}$ is the mean level spacing in the left/right quantum dot,
\begin{equation}
g=2\sum_k T_k
\end{equation}
is the dimensionless conductance of the barrier and
\begin{equation}
\beta=\sum_kT_k(1-T_k)/\sum_k T_k
\end{equation}
is the corresponding Fano factor.

Likewise, the WL correction to the $n-$th barrier conductance in
1d array of $N-1$ quantum dots with mean level spacings $\delta_n$
connected by $N$ barriers with dimensionless conductances $g_n$
and Fano factors $\beta_n$ reads
\begin{eqnarray}
\delta G^{WL}_n&=&-\frac{e^2 g_n}{4\pi^2 }\int_{0}^\infty dt
\big\{ \beta_n\big[\delta_n C_{n-1,n}(t)
\nonumber\\ &&
+\,\delta_{n-1} C_{n,n-1}(t)\big]+(1-\beta_n)\big[\delta_n C_{nn}(t)
\nonumber\\ &&
+\,\delta_{n-1}C_{n-1,n-1}(t)\big]\big\}.
\label{GWLn}
\end{eqnarray}

So far we discussed the local properties, namely WL corrections to
the conductivity tensor, $\delta \sigma^{WL}_{\alpha,\beta}(\bm{r},\bm{r}'),$ and
to the conductance of a single barrier, $\delta G^{WL}_{LR}$. Our main goal
is, however, to evaluate the WL correction to the conductance of the whole
system. For bulk
metals one finds that at large scales the WL correction (\ref{sigma1}) is local,
$\delta \sigma^{WL}_{\alpha,\beta}(\bm{r},\bm{r}')\propto \delta(\bm{r}-\bm{r}').$
In general though, there can exist other, non-local, contributions to the
conductivity tensor \cite{Kane}. Without going into details here, we only
point out that, even if these non-local terms are present,
one can still apply the standard Ohm's law arguments in order to obtain
the conductance of the whole sample. Specifically, in the case of 1d arrays one finds \cite{GZ06} (see also \cite{Argaman1})
\begin{eqnarray}
\delta G^{WL}&=&\frac{1}{\sum_{n=1}^N (G_n+\delta G^{WL}_n)^{-1}} -
\frac{1}{\sum_{n=1}^N G_n^{-1}}
\nonumber\\
&=& \frac{\sum_{n=1}^N \delta G^{WL}_n/g_n^2}{\left(\sum_{n=1}^N 1/g_n\right)^2} +
{\rm higher\; order\; terms}.\hspace{0.5cm}
\label{Garray}
\end{eqnarray}
Eqs. (\ref{GWL}), (\ref{GWLn}) and (\ref{Garray})  will be used to evaluate WL corrections for different configurations of quantum dots considered below.

\subsection{Examples}

\subsubsection{Single quantum dot}

\begin{figure}
\centerline{\includegraphics[width=5cm]{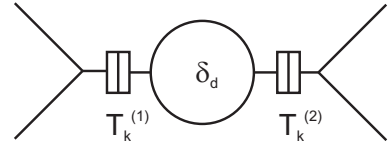}} \caption{Single
quantum dot connected to the leads via two barriers.}
\end{figure}
We start from the simplest case of a single quantum dot depicted in
Fig. 2. In this case the solution of Eq. (\ref{diff})
reads
\begin{eqnarray}
C_{11}(t)=\exp\left[-\frac{t}{\tau_D}-\frac{t}{\tau_H}-\frac{t}{\tau_\varphi}\right],
\end{eqnarray}
where $\tau_D=4\pi / (g_1+g_2)\delta_d$ is the dwell time,
and $\delta_d$ is the
mean level spacing in the quantum dot. All other components of
the Cooperon are equal to zero. From Eq. (\ref{GWL}) we get
\begin{eqnarray}
\delta G^{WL}_1&=&-\frac{e^2 g_1(1-\beta_1)\delta_d}{4\pi^2}\frac{1}{1/\tau_D+1/\tau_{H}+1/\tau_\varphi},
\nonumber\\
\delta G^{WL}_2&=&-\frac{e^2 g_2(1-\beta_2)\delta_d}{4\pi^2}\frac{1}{1/\tau_D+1/\tau_{H}+1/\tau_\varphi}.
\end{eqnarray}
According to Eq. (\ref{Garray}) the total WL correction
becomes
\begin{eqnarray}
\delta G^{WL}=-\frac{e^2\delta}{4\pi^2}
\frac{g_1g_2^2(1-\beta_1)+g_1^2g_2(1-\beta_2)}{(g_1+g_2)^2
\left({1}/{\tau_D}+{1}/{\tau_\varphi}+{1}/{\tau_H}\right)}.
\label{magres1qd}
\end{eqnarray}
Since $1/\tau_H\propto H^2$, the magnetoconductance has the
Lorentzian shape\cite{B}. In the limit $H=0$ and in the absence of
interactions ($\tau_\varphi\to\infty$) Eq. (\ref{magres1qd})
reduces to \cite{BB}
\begin{eqnarray}
\delta G^{WL}=-\frac{e^2}{\pi}
\frac{g_1g_2^2(1-\beta_1)+g_1^2g_2(1-\beta_2)}{(g_1+g_2)^3}.
\label{1qd}
\end{eqnarray}
As one can see for the case of low transmissions (for example in case of tunneling barriers) the WL corrections equals to zero.

\subsubsection{Two quantum dots}

\begin{figure}
\centerline{\includegraphics[width=3cm]{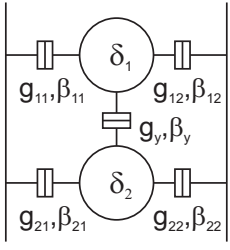}}
\caption{Most general system with two quantum dots}
\label{twodotgen}
\end{figure}

Next we consider the most general setup composed of two quantum
dots with the corresponding conductances and Fano factors defined
as in Fig. \ref{twodotgen}. The Cooperon is represented as a
$2\times 2$ matrix which zero frequency component satisfies the
following equation
\begin{eqnarray}
\left(\begin{array}{c}
g_{11}+g_{12}+g_y+\gamma_1 \hspace{1cm} -g_y \\
-g_y \hspace{1cm} g_{21}+g_{22}+g_y+\gamma_2
\end{array}\right)
\left(
\begin{array}{cc}
C_{11} & C_{12} \\
C_{21} & C_{22}
\end{array}\right)
\nonumber\\
=\,\left(
\begin{array}{cc}
4\pi/\delta_1 & 0 \\
0 & 4\pi/\delta_2
\end{array}
\right),
\end{eqnarray}
where
\begin{eqnarray}
\gamma_{1,2}=\frac{4\pi}{\delta_{1,2}}
\left(\frac{1}{\tau_{H1,2}}+\frac{1}{\tau_{\varphi 1,2}}\right).
\label{gamma}
\end{eqnarray}
Defining $\Delta=(g_{11}+g_{12}+g_y+\gamma_1)(g_{21}+g_{22}+g_y+\gamma_2)-g_y^2$, we get
\begin{eqnarray}
\left(
\begin{array}{cc}
C_{11} & C_{12} \\
C_{21} & C_{22}
\end{array}\right)=
\frac{4\pi}{\Delta}\left(
\begin{array}{c}
(g_{21}+g_{22}+g_y+\gamma_2)/\delta_1 \hspace{0.2cm} g_y/\delta_2 \\
g_y/\delta_1 \hspace{0.2cm} (g_{11}+g_{12}+g_y+\gamma_1)/\delta_2
\end{array}
\right).
\nonumber
\end{eqnarray}
With the aid of Eq. (\ref{GWL}) we can derive WL corrections for all
five barriers in our setup which we do not specify here for the sake of brevity (see \cite{GZ06} for further details).
\begin{figure}
\centerline{\includegraphics[width=8cm]{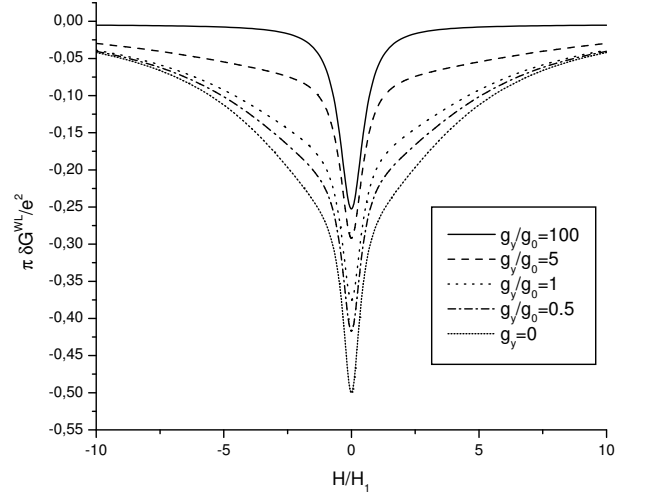}}
\caption{The magnetoconductance of two dots of Fig. \ref{twodotgen}
for $d_1,d_2\gg l_e,$ $d_1/d_2=5$, $g_{ij}=g_0$, $\beta_{ij}=0$, $\beta_y=0$, $\tau_{\varphi 1}=\tau_{\varphi 2}=\infty$.
Here $H_1=1/4\sqrt{\alpha_1\tau_{D1}}$ is the field at which weak localization is
effectively suppressed in the first dot.
For $g_y=0$ the magnetoconductance is given by superposition of two
Lorentzians with different widths (decoupled dots), while
for large $g_y$ only one Lorentzian survives corresponding to the contribution
of a one ``composite dot''.}
\label{cond2dots}
\end{figure}

WL correction to the conductance of the whole structure $\delta
G^{WL}$ is obtained from the general expression for the
conductance determined by Ohm's law:
\begin{eqnarray}
G&=&\big[G_{11}G_{12}(G_{21}+G_{22})+G_{21}G_{22}(G_{11}+G_{12})
\nonumber\\ &&
+\,G_y(G_{12}+G_{22})(G_{11}+G_{21})\big]
\nonumber\\ &&
\big/\,\big[ (G_{11}+G_{12})(G_{21}+G_{22})
\nonumber\\ &&
+\,G_y(G_{11}+G_{12}+G_{21}+G_{22}) \big].
\label{GOhm}
\end{eqnarray}
Substituting $G_{ij}\to G_{ij}+\delta G_{ij}^{WL}$ into this
formula and expanding the result to the first order in $\delta
G_{ij}^{WL}$, we get
\begin{eqnarray}
 \delta G^{WL}=\sum_{i,j=1,2}\frac{\partial G}{\partial  G_{ij}}\delta G_{ij}^{WL}
+\frac{\partial G}{\partial G_{y}}\delta G_{y}^{WL}.
\label{dwl2}
\end{eqnarray}
This general result for the WL correction to the conductance is illustrated in Fig. \ref{cond2dots} for a particular choice of the system parameters.

\begin{figure}
\centerline{\includegraphics[width=7.5cm]{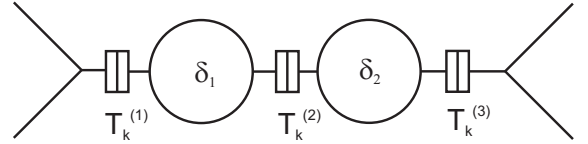}}
\caption{Two quantum dots in series.}
\label{twodot1}
\end{figure}

Of particular importance for us here is the system of two quantum
dots connected in series, as shown in Fig. \ref{twodot1}, i.e. in
the general structure of Fig. \ref{twodotgen} we set $G_{12}=G_{21}=0,$
$G_{11}=G_1,$ $G_y=G_2$, $G_{22}=G_3$, $\beta_{11}=\beta_1$,
$\beta_y=\beta_2$ and $\beta_{22}=\beta_3$. We also assume $H=0$
and $\tau_\varphi=\infty.$ WL corrections to the barrier
conductances then take the form
\begin{eqnarray}
\delta
G_1^{WL}&=&-\frac{e^2}{\pi}\frac{g_1(g_2+g_3)(1-\beta_1)}{g_1g_2+g_2g_3+g_1g_3},
\nonumber\\
\delta
G_2^{WL}&=&-\frac{e^2}{\pi}\frac{g_2(g_1+g_3)(1-\beta_2)+2g_2^2}{g_1g_2+g_2g_3+g_1g_3},
\nonumber\\
\delta
G_3^{WL}&=&-\frac{e^2}{\pi}\frac{g_3(g_1+g_2)(1-\beta_3)}{g_1g_2+g_2g_3+g_1g_3},
\end{eqnarray}
while Eq. (\ref{GOhm}) reduces to
\begin{eqnarray}
G&=&\frac{G_1G_{2}G_{3}} { G_{1}G_{2} +G_1G_{3}+G_{2}G_3}.
\label{GOhm2}
\end{eqnarray}
WL correction for the whole system then reads
\begin{eqnarray}
\delta G^{WL}&=&-\frac{e^2}{\pi}\frac{g_1g_2^2g_3^2(g_2+g_3)(1-\beta_1)}{(g_1g_2+g_2g_3+g_1g_3)^3}
\nonumber\\ &&
-\,\frac{e^2}{\pi}\frac{g_1^2g_2g_3^2(g_1+g_3)(1-\beta_2)}{(g_1g_2+g_2g_3+g_1g_3)^3}
\nonumber\\ &&
-\, \frac{e^2}{\pi}\frac{g_1^2g_2^2g_3(g_1+g_2)(1-\beta_3)}{(g_1g_2+g_2g_3+g_1g_3)^3}
\nonumber\\ &&
-\, \frac{2e^2}{\pi}\frac{g_1^2g_2^2g_3^2}{(g_1g_2+g_2g_3+g_1g_3)^3}.
\label{2qd}
\end{eqnarray}
In the limit of open quantum dots, i.e. $\beta_{1,2,3}=0$, we
reproduce the result \cite{Argaman1}. It is easy to see that
provided the conductance of one of the barriers strongly exceeds
two others, Eq. (\ref{2qd}) reduces to Eq. (\ref{1qd}). If all
three barriers are tunnel junctions, $\beta_{1,2,3} \to 1$, the
first three contributions in Eq. (\ref{2qd}) vanish, and only the
last contribution -- independent of the Fano factors -- survives
in this limit. If, on top of that, one of the tunnel junctions,
e.g. the central one, is less transparent than two others, $g_2
\ll g_1,g_3$, the result acquires a particularly simple (non-Lorentzian) form
\begin{eqnarray}
\delta G^{WL}=
- \frac{2e^2}{\pi}\frac{g_2^2}{\left(g_1+\gamma_1\right)\left(g_3+\gamma_2\right)},
\end{eqnarray}
with $\gamma_{1,2}$ defined in Eq. (\ref{gamma}).
Note that  $\delta G^{WL}\propto g_2^2$, i.e. this result is dominated by the
second order tunneling processes across the second barrier.

\subsubsection{1D array of identical quantum dots}

Let us now turn to 1d arrays of quantum dots depicted in
Fig. \ref{array1}. For simplicity, we will assume that our array
consists of $N-1$ identical quantum dots with the same level
spacing $\delta_n\equiv \delta_d$ and of $N$ identical barriers
with the same dimensionless conductance $g_n\equiv g$ and the same
Fano factor $\beta_n\equiv \beta$. We will also assume that the
quantum dots have the same shape and size so that
$\tau_{Hn}\equiv \tau_H$ and $\tau_{\varphi n}\equiv\tau_\varphi$.
For this system the Cooperon can also
be found exactly. The result reads
\begin{eqnarray}
C_{nm}(\omega)=\frac{2}{N}\sum_{q=1}^{N-1}\frac{\sin\frac{\pi qn}{N}\sin\frac{\pi qm}{N}}
{-i\omega+\frac{1}{\tau_H}+\frac{1}{\tau_\varphi}+\frac{1-\cos\frac{\pi q}{N}}{\tau_D}}.
\end{eqnarray}
Here $\tau_D=2\pi /g\delta_d$ and $\tau_H=1/16\alpha H^2.$
The WL correction then takes the form
\begin{eqnarray}
\delta G^{WL}=-\frac{e^2g\delta_d}{2\pi^2N^2}\sum_{q=1}^{N-1}
\frac{\beta\cos\frac{\pi q}{N}+1-\beta}
{\frac{1}{\tau_H}+\frac{1}{\tau_\varphi}+\frac{1-\cos\frac{\pi
q}{N}}{\tau_D}}. \label{WLN}
\end{eqnarray}

The sum over $q$ can be handled exactly and yields
\begin{eqnarray}
&& \delta G^{WL}=-\frac{e^2}{\pi N^2}
\bigg[
\left(N\frac{1+u^{2N}}{1-u^{2N}}-\frac{1+u^2}{1-u^2}\right)
\nonumber\\ &&\times\,
\frac{\beta(1+u^2)+2(1-\beta)u}{1-u^2} -(N-1)\beta
\bigg],
\label{magresN}
\end{eqnarray}
where
\begin{eqnarray}
u=1+\frac{\tau_D}{\tau_H}+\frac{\tau_D}{\tau_\varphi}
-\sqrt{\left(1+\frac{\tau_D}{\tau_H}+\frac{\tau_D}{\tau_\varphi}\right)^2-1}.
\label{uuu}
\end{eqnarray} In the tunneling limit $\beta=1$ and for
$\tau_\varphi\to\infty$ our result defined in Eqs. (\ref{magresN})-(\ref{uuu})
becomes similar -- though not exactly
identical -- to the corresponding result \cite{CN}.

If $\tau_\varphi$ is long enough, namely $
{1}/{\tau_\varphi}\lesssim E_{\rm Th}, $ where $E_{\rm
Th}=\pi^2/2N^2\tau_D$ is the Thouless energy of the whole array,
in Eqs. (\ref{WLN})-(\ref{magresN}) it is sufficient to set
$\tau_\varphi=\infty$. In this case the magnetic field $H$
significantly suppresses WL correction provided $1/\tau_H\gtrsim
E_{\rm Th}$ or, equivalently, if
\begin{eqnarray}
H\gtrsim H_N,\;\; H_N=\frac{1}{8N}\sqrt{\frac{\pi
g\delta_d}{\alpha}}. \label{HN}
\end{eqnarray}

\begin{figure}
\centerline{\includegraphics[width=8cm]{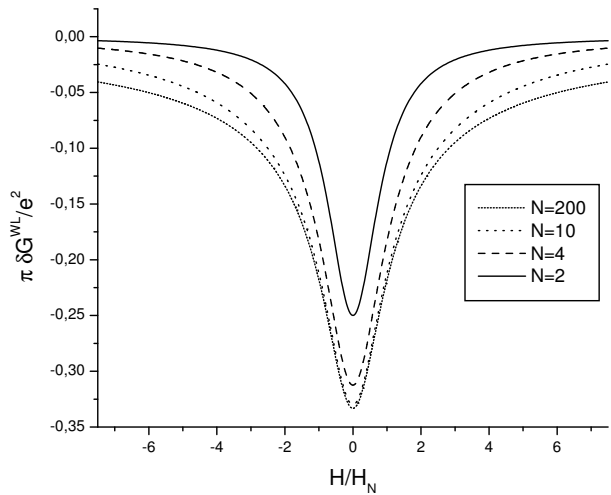}}
\caption{Magnetoconductance of a 1d array of $N-1$ identical open
($\beta=0$) quantum dots in the absence of interactions
($\tau_\varphi\to\infty$). The field $H_N$ is defined in Eq.
(\ref{HN}). } \label{twodot}
\end{figure}

In the opposite limit $1/\tau_\varphi\gtrsim E_{\rm Th}$
we find
\begin{eqnarray}
\delta G^{WL}=-\frac{e^2}{\pi N}\left[
\frac{\beta\left(1+\frac{\tau_D}{\tau_H}+\frac{\tau_D}{\tau_\varphi}\right)+1-\beta}
{\sqrt{\left(1+\frac{\tau_D}{\tau_H}+\frac{\tau_D}{\tau_\varphi}\right)^2-1}}
-\beta\right].
\end{eqnarray}
In particular, in the diffusive limit
$\tau_H,\tau_\varphi\gg\tau_D$ we get
\begin{eqnarray}
\delta G^{WL}=-\frac{e^2}{\pi
Nd}\sqrt{\frac{D\tau_H\tau_\varphi}{\tau_H+\tau_\varphi}},
\label{diffmet}
\end{eqnarray}
where we introduced the diffusion coefficient
\begin{eqnarray}
D=d^2/2\tau_D.
\label{D}
\end{eqnarray}
Eq. (\ref{diffmet}) coincides with the standard result for
quasi-1d diffusive metallic wire. Note, however, that the values
of $\tau_H$ within our model may differ from those for a metallic
wire. The ratio of the former to the latter is $\tau_H^{\rm
qd}/\tau_H^{\rm met}\sim \tau_{\rm fl}/\tau_D,$ where $\tau_{\rm
fl}\sim d/v_F$ is the flight time through the quantum dot. Since
typically $\tau_{\rm fl}<\tau_D$ we conclude that for the same
value of $D$ the magnetic field dephases electrons stronger in the
case of an array of quantum dots.

For a single quantum dot ($N=2$) Eq. (\ref{magresN})
reduces to
\begin{eqnarray}
\delta G^{WL}=-\frac{e^2(1-\beta)}{4\pi}
\frac{1}{\left(1+\frac{\tau_D}{\tau_H}+\frac{\tau_D}{\tau_\varphi}\right)}
\end{eqnarray}
in agreement with Eq. (\ref{magres1qd}).

For two identical quantum dots in series we obtain
\begin{eqnarray}
\delta G^{WL}=-\frac{e^2}{9\pi}\bigg[
\frac{2-\beta}{1+\frac{2\tau_D}{\tau_H}+\frac{2\tau_D}{\tau_\varphi}}
+\frac{\frac{2}{3}-\beta}{1+\frac{2\tau_D}{3\tau_H}+\frac{2\tau_D}{3\tau_\varphi}}
\bigg],
\end{eqnarray}
i.e. the magnetoconductance is just the sum of two Lorentzians in
this case.

Finally, in the absence of any interactions ($\tau_\varphi=\infty$) and
at $H=0$ we obtain
\begin{eqnarray}
\delta G^{WL}=-\frac{e^2}{\pi}\left[\frac{1}{3}-\frac{\beta}{N}
+\frac{1}{N^2}\left(\beta-\frac{1}{3}\right)\right].
\label{GWLarray}
\end{eqnarray}
In the limit $N\to \infty$ this result reduces to the standard one
for a long quasi-1d diffusive wire \cite{Mello} while for any finite $N$ we
reproduce the results for tunnel barriers \cite{CN} ($\beta \to 1$) and
open quantum dots \cite{Argaman2} ($\beta \to 0$).

The magnetoconductance of a 1d array of $N-1$ identical open
quantum dots in the absence of interactions is also illustrated in Fig. 6.

\section{Quantum decoherence by electron-electron interactions}

\subsection{Qualitative arguments}

Let us now include electron-electron interactions and analyze their impact on loss of phase coherence of electrons' wave functions. Before turning to a detailed calculation it is instructive to
discuss a simple qualitative picture demonstrating under which
conditions decoherence by electron-electron interactions is expected to occur.

\begin{figure}
\includegraphics[width=5.5cm]{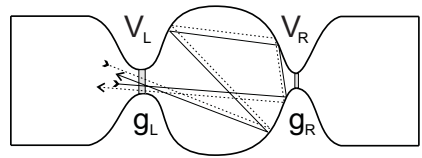}
\caption{Single quantum dot and a pair of time-reversed electron
paths. Fluctuating voltages $V_L$ and $V_R$ are assumed to drop
only across the barriers and not inside the dot.}
\end{figure}

Consider first the simplest system of two scatterers separated by
a cavity (quantum dot, Fig. 7) The WL correction to conductance of
a disordered system $G_{WL}$ is known to arise from interference
of pairs of time-reversed electron paths \cite{CS}. In the absence
of interactions for a single quantum dot of Fig. 7 this correction
was evaluated in the previous sections (see Eq. (\ref{1qd})).
The effect of
electron-electron interactions can be described in terms of
fluctuating voltages. Let us assume that the voltage can drop only
across the barriers and consider two time-reversed electron paths
which cross the left barrier (with fluctuating voltage $V_L(t)$)
twice at times $t_i$ and $t_f$, as it is shown in Fig. 7. It is easy to
see that the voltage-dependent random phase factor $\exp
(i\int_{t_i}^{t_f}V_L(t)dt)$ acquired by the electron wave
function $\Psi$ along any path turns out to be exactly the same as
that for its time-reversed counterpart. Hence, in the product
$\Psi\Psi^*$ these random phases cancel each other and quantum
coherence of electrons remains fully preserved. This implies that
for the system of Fig. 9 fluctuating voltages (which can mediate
electron-electron interactions) {\it do not cause any dephasing}.

This qualitative conclusion can be verified by means of more
rigorous considerations. For instance, it was demonstrated
\cite{GZ041} that the scattering matrix of the system remains
unitary in the presence of electron-electron interactions, which
implies that the only effect of such interactions is transmission
renormalization but not electron decoherence. A similar conclusion was reached \cite{Brouwer} by directly
evaluating the WL correction to the system conductance. Thus, for
the system of two scatterers of Fig. 7 electron-electron
interactions can only yield energy dependent (logarithmic at
sufficiently low energies) renormalization of the dot channel
transmissions \cite{GZ041,BN} but not electron dephasing.

\begin{figure}
\includegraphics[width=7.5cm]{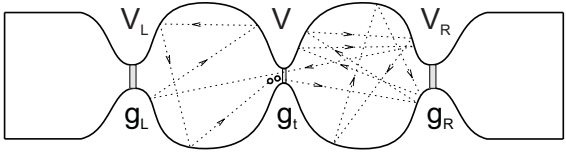}
\caption{Two quantum dots and a typical electron path.  Fluctuating
voltages $V_L$, $V$ and $V_R$ are again assumed to drop only
across the barriers.}
\end{figure}

Let us now add one more scatterer and consider the system of two
quantum dots depicted in Fig. 8. We again assume that fluctuating
voltages are concentrated at the barriers and not inside the
cavities. The phase factor accumulated along the path (see Fig. 8)
which crosses the central barrier twice (at times $t_i$ and
$t>t_i$) and returns to the initial point (at a time $t_f$) is
$e^{i[\varphi (t_i)-\varphi (t)]}$, where $\dot\varphi /e=V(t)$ is
the fluctuating voltage across the central barrier. Similarly, the
phase factor picked up along the time-reversed path reads
$e^{i[\varphi (t_f+t_i-t)-\varphi (t_f)]}$. Hence, the overall
phase factor acquired by the product $\Psi\Psi^*$ for a pair of
time-reversed paths is $\exp (i \Phi_{\rm tot})$, where
$$
\Phi_{\rm tot}(t_i,t_f,t)=\varphi (t_i)-\varphi
(t)-\varphi^+(t_f+t_i-t)+\varphi (t_f).
$$
Averaging over phase fluctuations, which for simplicity are
assumed Gaussian, we obtain
\begin{eqnarray}
&&\left\langle e^{i \Phi_{tot}(t_i,t_f,t) } \right\rangle
=\,e^{-\frac{1}{2} \left\langle \Phi_{tot}^2(t_i,t_f,t)
\right\rangle} \nonumber\\ &&
=\,e^{-2F(t-t_i)-2F(t_f-t)+F(t_f-t_i)+F(t_f+t_i-2t)},
\label{phase}
\end{eqnarray}
where we defined the phase correlation function
\begin{equation}
F(t)=\langle (\varphi(t)-\varphi(0))^2\rangle /2. \label{F}
\end{equation}
Should this function grow with time the electron phase coherence
decays and, hence, $G_{WL}$ has to be suppressed below its
non-interacting value due to interaction-induced electron
decoherence.

The above arguments are, of course, not specific to systems with
three barriers only. They can also be applied to any system with
larger number of scatterers, i.e. virtually to any disordered
conductor where -- exactly for the same reasons -- one also
expects non-vanishing interaction-induced electron decoherence at
any temperature including $T=0$. Below we will
develop a quantitative theory which will confirm and extend our
qualitative physical picture. We are going to give a complete
quantum mechanical analysis of the problem which fully accounts
for Fermi statistics of electrons and treats electron-electron
interactions in terms quantum fields produced internally by
fluctuating electrons.

\subsection{Nanorings with two quantum dots}

\subsubsection{The model and basic formalism}
\begin{figure}
\includegraphics[width=7.5cm]{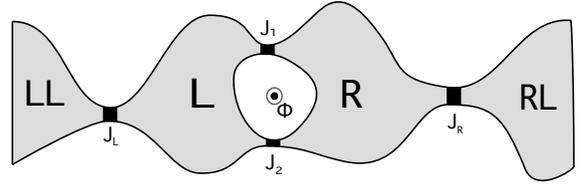}
\caption{Two quantum dots with magnetic flux.}
\end{figure}
Consider the system depicted in Fig. 9.
The structure consists of two chaotic quantum dots (L and R)
characterized by mean level spacing $\delta_L$ and $\delta_R$
which are the lowest energy parameters in our problem. These
(metallic) dots are interconnected via two tunnel junctions J$_1$
and J$_2$ with conductances $G_{t1}$ and $G_{t2}$ forming a
ring-shaped configuration as shown in Fig. 9. The left and right
dots are also connected to the leads (LL and RL) respectively via
the barriers J$_L$ and J$_R$ with conductances $G_L$ and $G_R$. We
also define the corresponding dimensionless conductances of all
four barriers as $g_{t1,2}=G_{t1,2}R_q$ and $g_{L,R}=G_{t1,2}R_q$,
where $R_q=2\pi /e^2$ is the quantum resistance unit.

The whole structure is pierced by the magnetic
flux $\Phi$ through the hole between two central barriers in such
way that electrons passing from left to right through different
junctions acquire different geometric phases. Applying a voltage
across the system one induces the current which shows AB
oscillations with changing the external flux $\Phi$. Note that
in the absence of the magnetic flux the system just reduces to that of
two connected in series quantum dots (cf. Fig. 5) which is also subject to weak localization effects. Thus, the model considered here allows to
analyze WL and AB effects within the same formalism to be developed below.
The system depicted in Fig. 9 is described by the effective
Hamiltonian:
\begin{eqnarray}
\hat H&=&\sum_{i,j=L,R}\frac{C_{ij}\hat {\bm V}_i\hat {\bm
V}_j}{2} +\hat {\bm H}_{LL}+\hat {\bm H}_{RL} \nonumber\\ &&
+\sum_{j=L,R}\hat {\bm H}_{j} +\hat{\bm T}_L +\hat {\bm T}_R+ \hat
{\bm T}, \label{H}
\end{eqnarray}
where $C_{ij}$ is the capacitance matrix, $\hat {\bm V}_{L(R)}$ is
the electric potential operator on the left (right) quantum dot,
$$
\hat {\bm
H}_{LL}=\sum\limits_{\alpha=\uparrow,\downarrow}\int\limits_{LL}d^3{\bm
r} \hat\Psi^\dagger_{\alpha,LL}({\bm r})(\hat
H_{LL}-eV_{LL})\hat\Psi_{\alpha,LL}({\bm r}),
 $$
$$
\hat {\bm
H}_{RL}=\sum\limits_{\alpha=\uparrow,\downarrow}\int\limits_{RL}d^3{\bm
r} \hat\Psi^\dagger_{\alpha,RL}({\bm r})(\hat
H_{RL}-eV_{RL})\hat\Psi_{\alpha,RL}({\bm r})
 $$
are the Hamiltonians of the left and right
leads, $V_{LL,RL}$ are the electric potentials of the leads fixed
by the external voltage source,
$$
\hat {\bm
H}_{j}=\sum\limits_{\alpha=\uparrow,\downarrow}\int\limits_{j}d^3{\bm
r} \hat\Psi^\dagger_{\alpha,j}({\bm r})(\hat H_{j}-e\hat {\bm
V}_{j})\hat\Psi_{\alpha,j}({\bm r})
 $$
defines the Hamiltonians of the left ($j=L$) and
right ($j=R$) quantum dots and
$$\hat H_j=\frac{(\hat p_\mu-\frac e c A_\mu(r))^2}{2m}-\mu+U_j(r)$$
is the one-particle Hamiltonian of electron in $j$-th quantum dot
with disorder potential $U_j(r)$. Electron transfer between the
left and the right quantum dots will be described by the
Hamiltonian
$$
\hat{\bm T}=\sum_{\alpha=\uparrow,\downarrow}\int_{J_1+J_2} d^2{\bm
r}\, \big[t({\bm r})\hat\Psi^\dagger_{\alpha,L}({\bm r})
\hat\Psi_{\alpha,R}({\bm r})+{\rm c.c.}\big].
$$
The Hamiltonian
$\hat {\bm T}_{L(R)}$ describing electron transfer between the
left dot and the left lead (the right dot and the right lead) is
defined analogously.

The real time evolution of the
density matrix of our system is described by means of the standard equation
\begin{equation}
\hat \rho(t)=e^{-i\hat Ht}\hat\rho_0\,e^{i\hat Ht},
\end{equation}
where $\hat H$ is given by Eq. (\ref{H}). Let us express the
operators $e^{-i\hat Ht}$ and $e^{i\hat Ht}$ via path integrals
over the fluctuating electric potentials $V_j^{F,B}$ defined
respectively on the forward and backward parts of the Keldysh
contour:
\begin{eqnarray}
e^{-i\hat Ht}&=&\int  DV_j^F\; {\rm T}\,\exp\left\{-i\int_0^t
dt'\hat H\left[V_j^F(t')\right]\right\},
\nonumber\\
e^{i\hat Ht}&=&\int  DV_j^B\; \tilde{\rm T}\,\exp\left\{i\int_0^t
dt'\hat H\left[V_j^B(t')\right]\right\}.
\end{eqnarray}
Here ${\rm T}\,\exp$ ($\tilde {\rm T}\,\exp$) stands for the time
ordered (anti-ordered) exponent.

Let us define the effective action of our system
\begin{eqnarray}
iS[V^F,V^B]&=&\ln\left( {\rm tr} \left[ {\rm
T}\,\exp\left\{-i\int_0^t dt'\hat H\left[V_j^F(t')\right]\right\}
\right.\right. \nonumber\\ &&\times\, \left.\left. \hat\rho_0
\tilde{\rm T}\,\exp\left\{i\int_0^t dt'\hat
H\left[V_j^B(t')\right]\right\} \right]\right)
\end{eqnarray}
Integrating out the fermionic variables we rewrite the action in
the form
\begin{equation}
    iS=iS_C+iS_{ext}+2 \bf Tr\ln \left[\check  G^{-1} \right].
\label{ac1}
\end{equation}
Here $S_C$ is the standard term describing charging effects,
$S_{ext}$ accounts for an external circuit and
\begin{equation}
  {\bf \check G^{-1}}=\left(\begin{array}{cccc}
   \hat G^{-1}_{LL} & \hat T_L  & 0 & 0 \\
     \hat T^\dag_L & \hat G^{-1}_L & \hat T & 0 \\
     0 & \hat T^\dag & \hat G^{-1}_R & \hat T_R \\
     0 & 0 & \hat T^\dag_R & \hat G^{-1}_{RL}
\end{array}\right).
\end{equation}
is the inverse Green-Keldysh function of electrons propagating in
the fluctuating fields. Here each quantum dot as well as two leads
is represented by the 2x2 matrix in the Keldysh space:
\begin{equation}
   \hat G^{-1}_i=\left(\begin{array}{cc}
    i\partial_t-\hat H_i+eV^F_i & 0 \\
     0 & -i\partial_t+\hat H_i -eV^B_i
\end{array}
\right)
\end{equation}

\subsubsection{Effective action}

Let us expand the exact action $iS$ (\ref{ac1}) in powers of $\hat
T$. Keeping the terms up to the fourth order in the tunneling
amplitude, we obtain
 \begin{eqnarray}
     iS\approx iS_C+iS_{ext}
     +iS_L+iS_R-2{\bf tr}\left[\hat G_L\hat T \hat G_R\hat T^\dag\right]\nonumber\\
     -{\bf tr}\left[\hat G_L\hat T \hat G_R\hat T^\dag\hat G_L\hat T \hat G_R\hat
     T^\dag\right].
     \label{action1}
 \end{eqnarray}
Here $iS_{L,R}$ are the contributions of isolated dots, the terms
$\propto t^2$ yield the Ambegaokar-Eckern-Sch\"on (AES) action
\cite{SZ} $iS^{AES}$ described by the diagram in Fig. 10a, and the
fourth order terms $\propto t^4$ (diagrams in Fig. 10b,c) account for the weak localization correction to the system conductance \cite{GZ08,GZ07}.

\begin{figure}
 \centering
\includegraphics[width=2.5in]{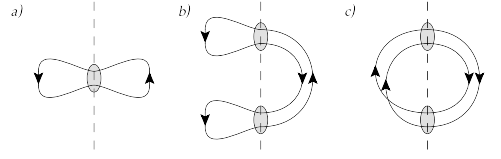}
\caption{\label{f2}Diagrammatic representation of different
contributions originating from expansion of the effective action
in powers of the central barrier transmissions: second order (AES)
terms (a) and different fourth order terms (b,c).}
\end{figure}

It is easy to demonstrate \cite{SGZ} that after disorder averaging
$iS^{AES}$ becomes independent of $\Phi$ and, hence, it does not
account for the AB effect investigated here.  After averaging the last
term in Eq. (\ref{action1}) over realizations of transmission
amplitudes and over disorder only the contribution generated by the diagram (c) keeps depending on the magnetic
flux and yields \cite{SGZ}
\begin{eqnarray}
iS^{WL}_{\Phi}=-\frac{ig_{t1}g_{t2}}{4\pi^2 N_L
N_R}\sum\limits_{m,n=1,2}e^{2i(\varphi_g^{(n)}-\varphi_g^{(m)})}\nonumber\\ \times\int
 d\tau_1 d\tau_2
\int dt_1 ...dt_4 C_L(\tau_1)C_R(\tau_2)\qquad\qquad\nonumber\\\times
e^{i(\varphi^+(t_2)-\varphi^+(t_3)+\varphi^+(t_4)-\varphi^+(t_1))}
\sin\frac{\varphi^-(t_1)}{2}\nonumber
\\ \times
\left[h(t_1-t_2-\tau_1)e^{i\frac{\varphi^-(t_2)}{2}}+\right.\qquad\qquad\qquad\qquad\nonumber\\\left.+f(t_1-t_2-\tau_1)e^{-i\frac{\varphi^-(t_2)}{2}}\right]\nonumber\\
\times
\left[h(t_2-t_3-\tau_2)e^{-i\frac{\varphi^-(t_3)}{2}}f(t_3-t_4+\tau_1)-\right.
\quad\nonumber\\ \left.-f(t_2-t_3-\tau_2)e^{i\frac{\varphi^-(t_3)}{2}}h(t_3-t_4+\tau_1)    \right]\nonumber\\
\times \left[e^{i\frac{\varphi^-(t_4)}{2}}f(t_4-t_1+\tau_2)+\right.\qquad\qquad\qquad\qquad\nonumber\\\left.+e^{-i\frac{\varphi^-(t_4)}{2}}h(t_4-t_1+\tau_2)\right]\nonumber\\
+\{L\leftrightarrow R,\varphi^{\pm}\rightarrow -\varphi^{\pm}\},
\label{sw2}
\end{eqnarray}
where $C_{L,R}(t)$ the Cooperons in the left and right dots,
$f(t)=\int f_F(E)dE/2\pi$ is the Fourier transform of the Fermi
function $f_F(E)$ and $h(t)=\delta(t)-f(t)$.
Here we also introduced the geometric phases
\begin{equation}
\varphi_{g}^{(1,2)}=\frac{e}{c}\int\limits_{L}^{R} dx_\mu
  A_\mu(x),
\end{equation}
where the integration contour starts in the left dot, crosses the
first ($\varphi_{g}^{(1)}$) or the second ($\varphi_{g}^{(2)}$)
junction and ends in the right dot. The difference between these
two geometric phases is $\varphi_g^{(1)}-\varphi_g^{(2)}=2\pi \Phi
/\Phi_0$. In addition, we defined the ``classical'' and the
``quantum'' components of the fluctuating phase
$\varphi^+(t)=(\varphi_F(t)+\varphi_B(t))/2$ and
$\varphi^-(t)=\varphi_F(t)-\varphi_B(t)$, where the phases $
\varphi_{F,B}(t)=e\int^t d\tau (V^{F,B}_R(\tau)-V^{F,B}_L(\tau))$
are defined on the forward and backward parts of the Keldysh
contour.

The above expression for the action $S^{WL}_{\Phi}$ (\ref{sw2})
fully accounts for coherent oscillations of the system conductance
in the lowest non-vanishing order in tunneling. The WL contribution to action of two quantum dots is recovered in exactly the same way \cite{GZ08}. The result is the similar except geometric phases should be omitted and the combination $g_{t1}g_{t2}$ should be substituted by $g_{t1}^2$ or $g_{t2}^2$.

\subsubsection{Aharonov-Bohm conductance and WL correction}

Let us now evaluate the current $I$ through our system. This
current can be split into two parts, $I=I_0+\delta I$, where $I_0$
is the flux-independent contribution and $\delta I$ is the quantum
correction to the current sensitive to the magnetic flux $\Phi$.
This correction is determined by the action $iS^{WL}_{\Phi}$, i.e.
\begin{equation}
 \delta I=-e\int\mathcal D^2\varphi^{\pm} \frac{\delta S^{WL}_{\Phi}[\varphi^+,\varphi^-]}{\delta \varphi^-(t)}
 e^{iS[\varphi^+,\varphi^-]}.
\label{IAB}
\end{equation}

In order to evaluate the path integral over the phases
$\varphi^{\pm}$ in (\ref{IAB}) we restrict our consideration to the most interesting for us metallic limit assuming that dimensionless
conductances $g_{L,R}$ are much larger than unity, while the
conductances $g_{t1}$ and $g_{t2}$ are small as compared to those
of the outer barriers, i.e.
\begin{equation}
g_L,g_R\gg 1,g_{t1},g_{t2}. \label{met}
\end{equation}
In the limit (\ref{met}) phase fluctuations can be considered
small down to exponentially low energies \cite{PZ91,Naz} in
which case it suffices to expand both contributions  up to the
second order $\varphi^{\pm}$. Moreover, this Gaussian
approximation becomes {\it exact} \cite{GGZ05,GZ01,GZ041,BN} in the
limit of fully open left and right barriers with $g_{L,R} \gg 1$.
Thus, in the metallic limit (\ref{met}) the integral (\ref{IAB})
remains Gaussian at all relevant energies and can easily be
performed.

This task can be accomplished with the aid of the following
correlation functions
\begin{equation}
   \langle\varphi^+(t)\rangle=eVt,\qquad
   \langle\varphi^-(t)\rangle=0,
\label{cf1}
\end{equation}
\begin{equation}
   \langle (\varphi^+(t)-\varphi^+(0))\varphi^+(0)\rangle=-F(t),
\label{cf2}
\end{equation}
\begin{equation}
   \langle \varphi^+(t)\varphi^-(0)+\varphi^-(t)\varphi^+(0)\rangle =2iK(|t|),
\label{cf3}
\end{equation}
\begin{equation}
   \langle \varphi^+(t)\varphi^-(0)-\varphi^-(t)\varphi^+(0)\rangle=2iK(t),
\label{cf4}
\end{equation}
\begin{equation}
   \langle \varphi^-(t)\varphi^-(0)\rangle=0,
\label{cf5}
\end{equation}
where the last relation follows directly from the causality
principle \cite{GZ}. Here and below we define $V=V_{RL}-V_{LL}$
to be the transport voltage across our system.

Note that the above correlation functions are well familiar from
the so-called $P(E)$-theory\cite{SZ,IN} describing electron
tunneling in the presence of an external environment which can
also mimic electron-electron interactions in metallic conductors.
They are expressed in terms of an effective impedance $Z(\omega)$
``seen'' by the central barriers J$_1$ and J$_2$
\begin{equation}
   F(t)=e^2\int\frac{d\omega}{2\pi}\coth\frac{\omega}{2T}\Re[Z(\omega)]\frac{1-\cos(\omega
   t)}{\omega},
\label{Ft}
\end{equation}
\begin{equation}
  K(t)=e^2\int\frac{d\omega}{2\pi}\Re[Z(\omega)]\frac{\sin(\omega
  t)}{\omega}.
\label{Kt}
\end{equation}
Further evaluation of these correlation functions for our system
is straightforward and yields
\begin{equation}
F(t)\simeq \frac{4}{g} \left(\ln\left|\frac{\sinh(\pi T t)}{\pi
 T\tau_{RC}}\right|+\gamma    \right),
\label{FFF}
\end{equation}
\begin{equation}
K(t)\simeq\frac{2\pi}{g}{\rm sign}(t), \label{KKK}
\end{equation}
where we defined $g=4\pi/e^2Z(0)$ and $\gamma\simeq0.577$ is the
Euler constant. Neglecting the contribution of external leads and
making use of the inequality (\ref{met}) we obtain $g\simeq
2g_Lg_R/(g_L+g_R)$. We observe that while $F(t)$ grows with time
at any temperature including $T=0$, the function $K(t)$ always
remains small and it can be safely ignored in the leading order in
$1/g \ll 1$. After that the Fermi function $f_F(E)$ drops out from
the final expression for the quantum correction to the current
\cite{GZ08,GZ07,SGZ}. Hence, the amplitude of AB oscillations is
affected by the electron-electron interaction only via the
correlation functions for the ``classical'' component of the
Hubbard-Stratonovich phase $\varphi^+$.

The expression for the current takes the form
\begin{equation}
  \delta I(\Phi )=-I_{AB}\cos(4\pi\Phi/\Phi_0)-I_{WL1}-I_{WL2},
\label{AB+q}
\end{equation}
where the first -- flux dependent -- term in the right-hand side
explicitly accounts for AB oscillations, while the terms
$I_{WL1,2}$ represent the remaining part of the quantum correction
to the current \cite{GZ08} which does not depend on $\Phi$.

Let us restrict our attention to the case of two identical quantum
dots with volume $\mathcal V$, dwell time $\tau_{D}$ and
dimensionless conductances $g_L=g_R \equiv g=4\pi/\delta\tau_D$,
where $\delta=1/ \mathcal V \nu$ is the dot mean level spacing and
$\nu$ is the electron density of states. In this case the
Cooperons take the form $C_L(t;{\bf x},{\bf y})=C_R(t;{\bf x},{\bf
y})=(\theta(t)/\mathcal V)e^{-t/\tau_D}$. We obtain \cite{SGZ}
\begin{eqnarray}
   I_{AB}=\frac{e^2g_{t1}g_{t2}\delta^2V}{4\pi^3}\int\limits_0^\infty d\tau_1 d\tau_2
   e^{-\frac{\tau_1+\tau_2}{\tau_D}-\mathcal F(\tau_1,\tau_2)}.
   \label{res}
\end{eqnarray}
\begin{eqnarray}
   I_{WL1,2}=\frac{e^2g_{t1,2}^2\delta^2V}{8\pi^3}\int\limits_0^\infty d\tau_1 d\tau_2
   e^{-\frac{\tau_1+\tau_2}{\tau_D}-\mathcal F(\tau_1,\tau_2)}.
\end{eqnarray}
where $\mathcal
F=2F(\tau_1)+2F(\tau_2)-F(\tau_1-\tau_2)-F(\tau_1+\tau_2)$.

In the absence of electron-electron interactions this formula
yields $I_{AB}^{(0)}=4e^2 g_{t1} g_{t2}V/(\pi g^2)$. In order to
account for the effect of interactions we substitute Eq.
(\ref{FFF}) into Eq. (\ref{res}). Performing time integrations at
high enough temperatures we obtain
\begin{equation}
 \frac{ I_{AB}}{I_{AB}^{(0)}}=\left\{\begin{array}{lc}
  e^{-\frac{8\gamma}{g}}\frac{(2\pi T\tau_{RC})^{8/g}}{1+4\pi T\tau_D/g},\quad &    \tau_D^{-1} \lesssim T \lesssim \tau_{RC}^{-1}, \\
 \frac{1}{2\tau_D}\left(\frac{g\tau_{RC}}{T}\right)^{1/2}, &  \tau_{RC}^{-1} \lesssim
 T,
  \end{array}\right.
\end{equation}
while in the low temperature limit we find
\begin{equation}
 \frac{I_{AB}}{I_{AB}^{(0)}}=e^{-\frac{8\gamma}{g}}\left(\frac{2\tau_{RC}}{\tau_D}\right)^{8/g}, \qquad T\lesssim
 \tau_D^{-1}.
 \end{equation}
Essentially the same results follow for $I_{WL1,2}$. These results demonstrate that interaction-induced suppression
of both AB oscillations and WL corrections in metallic dots with $\tau_{RC} \ll \tau_D$
persists down to $T=0$. The fundamental reason behind this
suppression is that the interaction of an electron with an
effective environment (produced by other electrons) effectively
breaks down the time-reversal symmetry and, hence, causes both
dissipation and dephasing for interacting electrons down to $T=0$
\cite{GZ}. In this respect it is also important to point out a
deep relation between interaction-induced electron decoherence and
the $P(E)$-theory \cite{SZ,IN} which was already emphasized
elsewhere \cite{GZ08,GZ07,SGZ}.

\subsection{Arrays of quantum dots and diffusive conductors}

One of the main conclusions reached above is
that the electron decoherence time is fully determined by
fluctuations of the phase fields $\varphi^+$ (and the correlation
function $F(t)$), whereas the phases $\varphi^-$ (and the response
function $K(t)$) are irrelevant for $\tau_\varphi$ causing only a
weak Coulomb correction to $G_{WL}$. This conclusion is general
being independent of a number of scatterers in our system. Note
that exactly the same conclusion was already reached in the case
of diffusive metals by means of a different approach
\cite{GZ}. Thus, in order to evaluate the decoherence
time for interacting electrons in arrays of quantum dots it is
sufficient to account for the fluctuating fields $V^+$ totally
ignoring the fields $V^-$. The corresponding calculation is
presented below.

\subsubsection{1d structures}

Let us consider a 1D array of $N-1$ quantum dots by $N$ identical
barriers as shown in Fig. 1. For simplicity, we will stick to the
case of identical barriers (with dimensionless conductance $g \gg
1$ and Fano factor $\beta$) and identical quantum dots (with mean
level spacing $\delta$ and dwell time $\tau_D=2\pi/\delta g$). The
WL correction to the system conductance has the form (see Eq. (\ref{GWLn})):
\begin{eqnarray}
 G_{WL}&=&-\frac{e^2g\delta}{4\pi^2N^2}\sum_{n=1}^N
\int_0^\infty dt\, \nonumber\\ &&\times\, \big\{
\beta\big[C_{n-1,n}(t)+C_{n,n-1}(t)\big] \nonumber\\ &&
+\,(1-\beta)\big[C_{nn}(t)+C_{n-1,n-1}(t)\big] \big\}. \label{GWLL}
\end{eqnarray}
The Cooperon $C_{nm}(t)$ is determined from a discrete version of
the diffusion equation. For non-interacting electrons and in the
absence of the magnetic field this equation reads
\begin{eqnarray}
\frac{\partial C_{nm}}{\partial
t}+\frac{2C_{nm}-C_{n-1,m}-C_{n+1,m}}{2\tau_D}=\delta_{nm}\delta(t).
\label{EqC}
\end{eqnarray}
The boundary conditions for this equation are $C_{nm}=0$ as long
as the index $n$ or $m$ belongs to one of the bulk electrode. The
solution of Eq. (\ref{EqC}) with these boundary conditions can
easily be obtained. We have
\begin{eqnarray}
C_{nm}^{(0)}(t)=\frac{2}{N}\sum_{q=1}^{N-1}\int\frac{d\omega}{2\pi}\,e^{-i\omega
t}\,\frac{\sin\frac{\pi qn}{N}\sin\frac{\pi qm}{N}}
{-i\omega+\frac{1-\cos\frac{\pi q}{N}}{\tau_D}}. \label{C0}
\end{eqnarray}
This solution can be represented in the form
$C^{(0)}_{nm}(t)=C^{\rm bulk}_{n-m}(t)-C^{\rm bulk}_{n+m}(t)$,
where
\begin{eqnarray}
C_{n-m}^{\rm
bulk}(t)=\frac{1}{N}\sum_{q=1}^{N-1}\int\frac{d\omega}{2\pi}\,e^{-i\omega
t}\,\frac{\cos\frac{\pi q(n-m)}{N}}
{-i\omega+\frac{1-\cos\frac{\pi q}{N}}{\tau_D}}. \label{Cbulk}
\end{eqnarray}

In the limit of large $N$ the term $C^{\rm bulk}_{n+m}(t)$ can be
safely ignored and we obtain $C_{nm}(t)\approx C^{\rm
bulk}_{n-m}(t)$. Let us express the contribution $C^{\rm
bulk}_{n-m}(t)$ as a sum over the integer valued paths
$\nu(\tau)$, which start in the $m-$th dot and end in the $n-$th
one (i.e. $\nu(0)=m$, $\nu(t)=n$) jumping from one dot to another
at times $t_j$. This expression  can be recovered if one expands
Eq. (\ref{Cbulk}) in powers of $\tau_D^{-1}\cos[{\pi q}/{N}]$ with
subsequent summation over $q$ in every order of this expansion.
Including additional phase factors acquired by electrons in the
presence of the fluctuating fields $V_\nu^+$, we obtain
\begin{eqnarray}
&&
C_{nm}(t)=\sum_{k=|n-m|}^\infty\;{\sum_{\nu(\tau)}}\bigg|_{\nu(0)=m}^{\nu(t)=n}
\nonumber\\ &&\times\, \frac{1}{(2\tau_D)^k}\int_0^t
dt_k\int_0^{t_k}dt_{k-1}\dots \int_0^{t_3}dt_2\int_0^{t_2}dt_1
\nonumber\\ &&\times\,
e^{-\frac{t-t_k}{\tau_D}}e^{-\frac{t_k-t_{k-1}}{\tau_D}}\dots
e^{-\frac{t_2-t_2}{\tau_D}}e^{-\frac{t_2-t_1}{\tau_D}}e^{-\frac{t_1}{\tau_D}}
\nonumber\\ &&\times\, \exp\left\{i\int_0^t
d\tau\big[eV^+_{\nu(\tau)}(\tau)-eV^+_{\nu(t-\tau)}(\tau)\big]\right\}.
\end{eqnarray}
Averaging over Gaussian fluctuations of voltages $V^+$ and
utilizing the symmetry of the voltage correlator $ \langle
V_{\nu_1}^+(\tau_1) V_{\nu_2}^+(\tau_2)\rangle =\langle
V_{\nu_2}^+(\tau_1) V_{\nu_1}^+(\tau_2)\rangle$, we get
\begin{eqnarray}
&&
C_{nm}(t)=\sum_{k=|n-m|}^\infty\;{\sum_{\nu(\tau)}}\bigg|_{\nu(0)=m}^{\nu(t)=n}
\nonumber\\ &&\times\, \frac{e^{-t/\tau_D}}{(2\tau_D)^k}\int_0^t
dt_k\int_0^{t_k}dt_{k-1}\dots \int_0^{t_3}dt_2\int_0^{t_2}dt_1
\nonumber\\ &&\times\, \exp\bigg\{-e^2\int_0^t d\tau_1\int_0^t
d\tau_2 \big[ \langle V_{\nu(\tau_1)}^+(\tau_1)
V_{\nu(\tau_2)}^+(\tau_2)\rangle \nonumber\\ && -\,\langle
V_{\nu(\tau_1)}^+(\tau_1) V_{\nu(t-\tau_2)}^+(\tau_2)\rangle \big]
\bigg\}. \label{C2}
\end{eqnarray}

The correlator of voltages can be derived with the aid of
the $\sigma$-model approach developed in Sec. 2 of this paper.
Integrating over Gaussian fluctuations of the $Q$-fields one
arrives at the quadratic action for the fluctuating fields $V^+$
which has the form
\begin{eqnarray}
&& iS=\frac{2i}{N}\sum_{q=1}^N\int\frac{d\omega}{2\pi}
\bigg[4C\left(1-\cos\frac{\pi q}{N}\right)+C_g \nonumber\\ &&
+\,\frac{g\tau_De^2}{\pi}\frac{1-\cos\frac{\pi q}{N}}
{-i\omega\tau_D+1-\cos\frac{\pi q}{N}}
\bigg]V^+_q(\omega)V^-_q(-\omega) \nonumber\\ &&
-\,\frac{2}{N}\sum_{q=1}^N\int\frac{d\omega}{2\pi}
\frac{g\tau_D^2e^2}{\pi} \frac{\left(1-\cos\frac{\pi
q}{N}\right)\omega\coth\frac{\omega}{2T}}
{\omega^2\tau_D^2+\left(1-\cos\frac{\pi q}{N}\right)^2}
\nonumber\\ &&\times\, V^-_q(\omega)V^-_q(-\omega). \label{S11}
\end{eqnarray}
Here we defined
\begin{equation}
V_q^\pm(\omega)=\sum_{n=1}^{N-1}\int dt\; \sin\frac{\pi q}{N}\,
e^{i\omega t}\, V_n^\pm(t).
\end{equation}
The action (\ref{S11}) determines the expressions for both
correlators $\langle V^+V^+\rangle$ ($F$-function) and $\langle
V^+V^-\rangle$ ($K$-function) responsible respectively for
decoherence and Coulomb blockade correction to WL. Since our aim
is to describe electron decoherence, only the first out of these
two correlation functions is of importance for us here. It reads
\begin{eqnarray}
&& \langle
V^+_n(t_1)V^+_m(t_2)\rangle=\frac{2}{N}\sum_{q=1}^{N-1}\int\frac{d\omega}{2\pi}\,
e^{-i\omega(t_1-t_2)} \nonumber\\ &&\times\, \frac{\frac{g
e^2}{\pi}\left(1-\cos\frac{\pi q}{N}\right)\, \sin\frac{\pi
qn}{N}\sin\frac{\pi qm}{N}} {\left| 4C\left(1-\cos\frac{\pi
q}{N}\right)+C_g +\frac{g\tau_De^2}{\pi}\frac{1-\cos\frac{\pi
q}{N}} {-i\omega\tau_D+1-\cos\frac{\pi q}{N}} \right|^2}
\nonumber\\ &&\times\, \frac{\tau_D^2\,\omega
\coth\frac{\omega}{2T}} {\omega^2\tau_D^2+\left( 1-\cos\frac{\pi
q}{N}\right)^2}. \label{VV}
\end{eqnarray}
In the continuous limit $N \gg 1$ and for sufficiently low
frequencies $\omega \ll 1/\tau_D$ both correlators $\langle
V^+V^+\rangle$ and $\langle V^+V^-\rangle$ defined by Eq.
(\ref{S11}) reduce to those of a diffusive metal \cite{GZ}.

To proceed let us consider diffusive paths $\nu(\tau)$, in which
case one has
\begin{eqnarray}
\langle V_{\nu(\tau_1)}^+(\tau_1) V_{\nu(\tau_2)}^+(\tau_2)\rangle
&\approx & \frac{1}{N-1}\sum_{n,m=1}^{N-1}\langle V_{n}^+(\tau_1)
V_{m}^+(\tau_2)\rangle \nonumber\\ &&\times\,
D_{nm}(|\tau_1-\tau_2|),
\label{VVD}
\end{eqnarray}
where $D_{nm}(\tau)$ is the diffuson. For $H \to 0$ it exactly
coincides with the Cooperon for non-interacting electrons
(\ref{C0}), $D_{nm}(t)=C_{n,m}^{(0)}(t)$, i.e.
\begin{eqnarray}
D_{nm}(t)=\frac{2}{N}\sum_{q=1}^{N-1}\int\frac{d\omega}{2\pi}\,e^{-i\omega
t}\,\frac{\sin\frac{\pi qn}{N}\sin\frac{\pi qm}{N}}
{-i\omega+\frac{1-\cos\frac{\pi q}{N}}{\tau_D}}. \label{DC0}
\end{eqnarray}
Substituting Eq. (\ref{VVD}) into (\ref{C2}), we obtain
\begin{equation}
C_{nm}(t)\approx C^{(0)}_{nm}(t)\, e^{-\mathcal{F}(t)}, \label{C3}
\end{equation}
where
\begin{eqnarray}
\mathcal{F}(t)&=&\frac{e^2}{N-1}\sum_{n,m=1}^{N-1}\int_{0}^t
dt_1dt_2 \langle V_n^+(t_1)V_m^+(t_2) \rangle \nonumber\\
&&\times\, \big[ D_{nm}(|t_1-t_2|)-D_{nm}(|t-t_1-t_2|)\big]
\label{F3}
\end{eqnarray}
is the function which controls the Cooperon decay  in time, i.e.
describes electron decoherence for our 1d array of quantum dots.
The WL correction $G_{WL}$ in the presence of electron-electron
interactions is recovered by substituting the result (\ref{C3})
into Eq. (\ref{GWLL}).

Since the behavior of the latter formula was already analyzed in
details earlier  there is no need to repeat this
analysis here. The dephasing time $\tau_\varphi$ can be extracted
from the equation $\mathcal{F}(\tau_\varphi)=1$. From Eq.
(\ref{F3}) with a good accuracy we obtain
\begin{eqnarray}
\frac{1}{\tau_\varphi}=\frac{e^2}{N-1}\sum_{n,m=1}^{N-1}\int d\tau
\langle V_n^+(\tau)V_m^+(0) \rangle D_{nm}(\tau). \label{tau11}
\end{eqnarray}
Combining this formula with Eqs. (\ref{VV}) and (\ref{DC0}), in
the most interesting limit $T\to 0$ and for $\tau_D\gg R(4C+C_g)$
we find
\begin{eqnarray}
\frac{1}{\tau_{\varphi 0}}&=&\frac{1}{2g\tau_D(N-1)}
\sum_{q=1}^{N-1}
\ln\frac{2e^2}{\delta\left(4C\left(1-\cos\frac{\pi
q}{N}\right)+C_g\right)}, \nonumber
\end{eqnarray}
which yields
\begin{equation}
\tau_{\varphi 0}=\frac{2g\tau_D}{\ln(4\tilde E_C/\delta)}
=\frac{4\pi}{\delta\ln(4\tilde E_C/\delta)}, \label{t1}
\end{equation}
where $\tilde E_C=e^2/2C_g$ for $C_g\gg C$ and  $\tilde
E_C=e^2/4C$ in the opposite case $C_g\ll C$.

In order to determine the dephasing length $L_\varphi
=\sqrt{D\tau_\varphi}$ let us define the diffusion coefficient
\begin{equation}
D=\frac{d^2}{2\tau_D}=\frac{d^2 g\delta}{4\pi}, \label{DDD}
\end{equation}
where $d\equiv \mathcal{V}^{1/3}$ is the average dot size.
Combining Eqs. (\ref{t1}) and (\ref{DDD}), at $T=0$ we obtain
\begin{equation}
L_{\varphi 0}= \sqrt{D\tau_{\varphi 0}}= d\sqrt{g/\ln (4\tilde
E_C/\delta )}.
\end{equation}

At non-zero $T$ thermal fluctuations provide an additional
contribution to the dephasing rate $1/\tau_\varphi$. Again
substituting Eqs. (\ref{VV}) and (\ref{DC0}) into (\ref{tau11}),
we get
\begin{equation}
\frac{1}{\tau_\varphi(T)}\simeq \frac{1}{\tau_{\varphi
0}}+\frac{\pi T}{3g}\min\{ N,N_\varphi \}, \label{tauT}
\end{equation}
where $N_\varphi =L_\varphi /d \sim \sqrt{\tau_\varphi/\tau_D}$ is
the number of quantum dots within the length $L_\varphi$. We
observe that for sufficiently small $N<N_\varphi$ (but still $N\gg
1$) the dephasing rate increases linearly both with temperature
and with the number $N$. At larger $N>\sqrt{g/\ln[{4\tilde
E_C}/{\delta}]}$ and/or at high enough temperatures $N_\varphi$
becomes smaller than $N$ and Eq. (\ref{tauT}) for $\tau_\varphi$
should be resolved self-consistently. In this case we obtain
\begin{equation}
\tau_\varphi\simeq (3g\sqrt{\tau_D}/\pi T)^{2/3} \label{AAK}
\end{equation}
thus reproducing the well known result \cite{AAK}. Eq. (\ref{tauT}) also
allows to estimate the temperature $ T^*\simeq 2\pi
g/[\tau_{\varphi 0}\min\{N,N_\varphi\}]$ at which the crossover to
the temperature-independent regime (\ref{t1}) occurs. We find
\begin{eqnarray}
&& T^*\simeq \frac{3\ln[{4\tilde E_C}/{\delta}]}{2\pi
N\tau_D},\;\;\; N\lesssim \sqrt{\frac{g}{\ln[{4\tilde
E_C}/{\delta}]}}, \nonumber\\
&& T^*\simeq \frac{3\ln^{3/2}[{4\tilde
E_C}/{\delta}]}{2\pi\tau_D\sqrt{g}},\;\;\;
 N\gtrsim \sqrt{\frac{g}{\ln[{4\tilde E_C}/{\delta}]}}.
\end{eqnarray}

\subsubsection{Good metals and granular conductors}

The above analysis and conclusions can be generalized further to
the case 2d and 3d structures. This generalization is absolutely
straightforward (see, e.g., \cite{GZ06}) and therefore is not
elaborated here. At $T \to 0$ one again arrives at the same result
for $\tau_{\varphi 0}$ (\ref{t1}).

Now we discuss the relation between our present results and those
derived earlier for weakly disordered metals by means of a
different approach \cite{GZ}. Let us express the dot
mean level spacing via the average dot size $d$ as $\delta=
1/N_0d^3$ (where $N_0=mp_F/2\pi^2$ is the electron density of
states at the Fermi level). Then we obtain
\begin{equation}
D=\frac{g}{4\pi N_0 d}.
\label{DDDD}
\end{equation}
Below we consider two different physical limits of $(a)$ good
metals and $(b)$ strongly disordered (granular) conductors. For
the model $(a)$ we assume that quantum dots are in a good contact
with each other. In this case $g$ scales linearly with the contact
area $\mathcal{A}=\gamma d^2$, where $\gamma$ is a numerical
factor of order (typically smaller than) one which particular
value depends on geometry. For weakly disordered metals most
conducting channels in such contacts can be considered open.
Hence, $g=p_F^2\mathcal{A}/2\pi$ and
\begin{equation}
D=\gamma v_Fd/4, \label{vFd}
\end{equation}
i.e. $D \propto d$. If most channels are not fully transparent,
then the factor $\gamma$ in (\ref{vFd}) also accounts for
their transmissions. Comparing Eq. (\ref{vFd})
with the standard definition of $D$ for a bulk diffusive
conductor, $D=v_Fl/3$, we immediately observe that within our
model the average dot size is comparable to the elastic mean free
path, $l \sim \gamma d$, as it should be for weakly disordered
metals.

Expressing $\tau_{\varphi 0}$ (\ref{t1}) via $D$, in this limit we
get
\begin{equation}
\tau_{\varphi 0}=\frac{64}{\pi
\gamma^3}\frac{m^2}{v_F^2}\frac{D^3}{\ln (D/D_{c1})}, \label{D3}
\end{equation}
where $m$ is the electron mass and $D_{c1}$ is constant which
depends on $\tilde E_C$. Estimating, e.g., $\tilde E_C \approx
e^2/2d$,  one obtains $D_{c1}^{-1}=4e\sqrt{2N_0}/\gamma v_F$.

Note that apart from an unimportant numerical pre-factor and the
logarithm in the denominator of Eq. (\ref{D3}) the latter result
for $\tau_{\varphi 0}$ coincides with that derived
for a bulk diffusive metal within the
framework of a completely different approach \cite{GZ}. Within that approach local properties of the model
remain somewhat ambiguous and, hence, in the corresponding
integrals in \cite{GZ} we could not avoid using an
effective high frequency cutoff procedure. This cutoff yields the
correct leading dependence $\tau_{\varphi 0} \propto D^3$ and it
only does not allow to recover an additional logarithmic
dependence on $D$ in (\ref{D3}). Our present approach is
divergence-free and, hence, it does not require any cutoffs.

We can also add that Eq. (\ref{t1}) also agrees with our earlier
results \cite{GZ} derived for quasi-1d and quasi-2d
metallic conductors. Provided the
transversal size $a$ of our array is
smaller than $d$ one should set $\mathcal{A} \sim da$ for 2d and
$\mathcal{A} \sim a^2$ for 1d conductors.
Then Eq. (\ref{t1}) yields  $\tau_{\varphi 0} \propto D^2/\ln D$
and  $\tau_{\varphi 0} \propto D/\ln D$  respectively in 2d and 1d cases.
Up to the factor $\ln D$ these dependencies coincide with ones derived
previously \cite{GZ}.

Now let us turn to the model $(b)$ of strongly disordered and/or
granular conductors. In contrast to the situation $(a)$, we will
assume that the contact between dots (grains) is rather poor, and
inter-grain electron transport may occur only via limited number
of conducting channels. In this case the average dimensionless
conductance $g$ can be approximated by some
$\mathcal{A}$-independent constant $g=g_c$. Substituting $g_c$
instead of $g$ into Eq. (\ref{DDDD}) we observe that in the case
of strongly disordered structures one can expect $D \propto 1/d$.
Accordingly, for $\tau_{\varphi 0}$ (\ref{t1}) one finds
\begin{equation}
\tau_{\varphi 0}=\frac{g_c^3}{32\pi^2 N_0^2D^3\ln (D_{c2}/D)},
\label{D-3}
\end{equation}
where $D_{c2}$ again depends on $\tilde E_C$. For $\tilde E_C
\approx e^2/2d$ we have $D_{c2}^{-1}=2\pi\sqrt{2N_0}/eg_c$. Hence,
the dependence of $\tau_{\varphi 0}$ on $D$ for strongly
disordered or granular conductors (\ref{D-3}) is {\ it
qualitatively} different from that for sufficiently clean metals
(\ref{D3}).

One can also roughly estimate the crossover between the regimes
$(a)$ and $(b)$ by requiring the values of $D=\gamma v_Fd/4$
(\ref{vFd}) and $D=g_c/4\pi N_0d$ to be of the same order. This
condition yields $(p_Fd)^2 \sim 2\pi g_c/\gamma$, and we arrive at
the estimate for $D$ at the crossover
\begin{equation}
D \approx \frac{0.6\hbar}{m} \sqrt{\frac{g_c}{\gamma}}.
\label{estD}
\end{equation}
Here we restored the Planck constant $\hbar$ set equal to unity
elsewhere in our paper.

\subsubsection{Ring composed of quantum dots}
\begin{figure}[t]
 \centering
\includegraphics[width=2.5in]{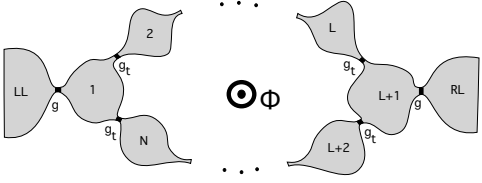}
\caption{\label{f5}Ring composed of $N$ quantum dots}
\end{figure}
Now let us turn to a ring-shaped nanostructure as shown in Fig. \ref{f5}. For simplicity we will consider the case of identical quantum dots
(with mean level spacing $\delta$ and dwell time
$\tau_D=2\pi/(g\delta)$) coupled by junctions with conductances
$g_t$ and the Fano-factor $\beta_t$. Leads are coupled to the ring
at the dots with numbers $1$ and $L+1$ by junctions with
conductance $g$. The interference correction to the conductance of
n-th junction $\delta G_n$ was already derived in Sec. 2 by means of the non-linear sigma-model approach. We obtain
\begin{eqnarray}
   \delta G_n=-\frac{e^2  g_t\delta}{4\pi^2}\int\limits_0^\infty dt
   [\beta_t C_{n,n+1}(t)e^{\frac{4\pi i\Phi }{N\Phi_0}}+\nonumber\\+(1-\beta_t)(C_{n,n}(t)+C_{n+1,n+1}(t))+\nonumber\\
   +\beta_tC_{n+1,n}(t)e^{-\frac{4\pi i\Phi }{N\Phi_0}}],
\label{dG1}
\end{eqnarray}
where $C_{m,n}(t)$ is the Cooperon. The
quantum correction to conductance of the whole system can be
obtained with the aid of the Kirchhoff's law. For the case $Ng\ll
g_t$ considered here one finds
\begin{equation}
\delta G=\frac{NL(N-L)g^2}{(2Ng_t+L(N-L)g)^2}\delta g\approx
\frac{L(N-L)g^2}{4Ng_t^2}\delta G_n.
\label{dG}
\end{equation}

Further procedure is analogous to that implemented above for 1d arrays. The main difference of the present ring-shaped geometry just concerns the form of diffusons $D_{mn}(t)$, cooperons $C_{mn}^{(0)}(t)$ and the fluctuating voltage correlators $F_{mn}(t)=\langle V^+_m(t)
V^+_n(0)\rangle_{V^+}$. We obtain
\begin{equation}
  D_{mn}(t)=\frac{\tau_D}{N}\sum\limits_{q=1}^N\int\frac{d\omega}{2\pi}
  \frac{e^{-i\omega t+\frac{2\pi i
  q}{N}(m-n)}}{-i\omega\tau_D+\epsilon(q)},
\end{equation}
\begin{equation}
  C_{mn}^{(0)}(t)=\frac{\tau_D}{N}\sum\limits_{q=1}^N
  \int\frac{d\omega}{2\pi}\frac{e^{-i\omega t+\frac{2\pi i
  q}{N}(m-n)}}{-i\omega\tau_D+\epsilon(q-2\Phi/\Phi_0)},
\end{equation}
and
\begin{equation}
 F_{mn}(t)=\frac{\tau_D}{N}\sum\limits_{q=1}^N\int\frac{d\omega}{2\pi}e^{-i\omega t}\omega\coth\frac{\omega}{2T}\frac{f(q)e^{\frac{2\pi i q}{N}(m-n)}}{\omega^2\tau_D^2+\varepsilon^2(q)},
\end{equation}
where
\begin{equation}
 f(q)=\frac{g_t\tau_D
 e^2}{\pi}\frac{\epsilon(q)}{(4C\epsilon(q)+C_g)^2},
\end{equation}
\begin{equation}
\varepsilon(q)=\epsilon(q)+\frac{g_t\tau_D e^2}{\pi}\frac{\epsilon(q)}{4C\epsilon(q)+C_g}
\end{equation}
and $\epsilon(q)=1-\cos\frac{2\pi q}{N}$. As above, here $C$ and
$C_g$ denote respectively the junction and the dot capacitances.

The above equations are sufficient to evaluate the function
$\mathcal F(t)$ in a general form. Here we are primarily
interested in AB oscillations and, hence, we only need to account
for the flux-dependent contributions determined by the electron
trajectories which fully encircle the ring at least once.
Obviously, one such traverse around the ring takes time $t\geq
N^2\tau_D$. Hence, the behavior of the function $\mathcal F(t)$
only at such time scales needs to be studied for our present
purposes. In this long time limit $\mathcal F(t)$ is a linear
function of time with the corresponding slope
\begin{eqnarray}
 \mathcal F'(t\geq N^2\tau_D)\approx\qquad\qquad\qquad\qquad\qquad\qquad\qquad\nonumber\\\approx\frac{2e^2\tau_D^2}{N}\sum\limits_{q=1}^{N-1}\int\frac{d\omega}{2\pi}\frac{f(q)\epsilon(q)\omega\coth\frac{\omega}{2T}}{(\omega^2\tau_D^2+\epsilon^2(q))(\omega^2\tau_D^2+\varepsilon^2(q))}
 \label{def}
\end{eqnarray}
This observation implies that at such time scales
electron-electron interactions yield exponential decay of the
Cooperon in time
\begin{equation}
C_{mn}(t)\approx
C_{mn}^{(0)}(t)e^{-\frac{t}{\tau_\phi}}
\end{equation}
where
\begin{equation}
\frac{1}{\tau_\phi}=\mathcal F'(t\geq N^2\tau_D)
\end{equation}
is the effective dephasing time for our problem. In the case
$C_g\gg C$ and $\tau_D\gg\tau_{RC}\equiv 2\pi C_g/(e^2 g_t)$ from
Eq. (\ref{dephtime}) we obtain
 \begin{equation}
    \frac{1}{\tau_\phi}=\left\{\begin{array}{cc}
    \frac{\delta}{\pi}\ln\frac{4E_C}{\delta} & \qquad\qquad T\ll 1/N\tau_D ,\\
       \frac{\pi N T }{3 g_t} & \qquad\qquad T\gg 1/N\tau_D,
    \end{array} \right.
\label{dephtime}
\end{equation} where $E_C=e^2/(2C_g)$. These
expressions are, of course, fully consistent with the results
 derived above in the case of 1d chains of quantum dots and
weakly disordered diffusive conductors, cf. also \cite{GZ}.

Let us emphasize again that the above results for $\mathcal F(t)$
apply at sufficiently long times which is appropriate in the case
of AB conductance oscillations. At the same time, other physical
quantities, such as, e.g., weak localization correction to
conductance can be determined by the function $\mathcal F(t)$ at
shorter time scales. Our general results allow to easily recover
the corresponding behavior as well. For instance, at $T\gg\tau_D$
and $t \ll N^2\tau_D$ we get
\begin{equation}
   \mathcal F(t)\approx \frac{4T }{3 g_t}\left(\frac{2\pi}{\tau_D}\right)^{1/2} t^{3/2}+...
\end{equation}
in agreement with the results \cite{GZ07}. This expression yields
the well known dependence $\tau_\phi \propto T^{-2/3}$ which -- in
contrast to Eq. (\ref{dephtime}) -- does not depend on $N$ and
remains applicable in the high temperature limit.

To proceed further let us integrate the expression for the
Cooperon over time. We obtain
\begin{eqnarray}
  \int\limits_0^{\infty}C_{mn}(t)dt=\qquad\qquad\qquad\qquad\qquad\qquad\nonumber\\
  =\frac{\tau_D}{N}\sum\limits_{q=1}^N\frac{e^{\frac{2\pi i q}{N}(m-n)}}{\epsilon(q-2\Phi/\Phi_0)+\tau_D/\tau_{\phi}+g/(g_t
  N)},
\label{tint}
\end{eqnarray}
where the term $g/(g_t N)$ in the denominator accounts for the
effect of external leads and remains applicable as long as $Ng\ll
g_t$. Combining Eqs. (\ref{dG1}), (\ref{dG}) and (\ref{tint})
after summation over $q$ we arrive at the final result
\begin{eqnarray}
\delta G^{AB}=\frac{e^2L(N-L)g^2}{2\pi Ng_t^2}\qquad\qquad\qquad\qquad\nonumber\\
\times\frac{(\beta_t\alpha+1-\beta_t)(z^{-N}-\cos(4\pi\Phi/\Phi_0))}{\sqrt{\alpha^2-1}(z^N+z^{-N}-2\cos(4\pi\Phi/\Phi_0))},
\label{final}
\end{eqnarray}
where $\alpha=1+\frac{\tau_D}{\tau_\phi}+\frac{g}{g_tN}$ and
$z=\alpha+\sqrt{\alpha^2-1}$.
This equation with Eq. (\ref{dephtime}) fully determines AB
oscillations of conductance in nanorings composed of metallic
quantum dots in the presence of electron-electron interactions.

Expanding Eq. (\ref{final}) in Fourier series we obtain
\begin{equation}
   \delta G^{AB}=\sum\limits_{k=1}^{\infty} \delta G^{(k)}\cos\left(4\pi k\Phi/\Phi_0\right)
\end{equation}
where
\begin{equation}
\delta G^{(k)}=-\frac{e^2L(N-L)g^2(\beta_t\alpha+1-\beta_t)}{2\pi N g_t^2\sqrt{\alpha^2-1}}z^{-N|k|}
\end{equation}
In the limit $\tau_\phi\gg\tau_D$ we have $ z\approx
1+\sqrt{2\tau_D/\tau_\phi}+...$, hence $\delta G^{(k)}$ behaves as
\begin{equation}
\delta G^{(k)}\propto e^{-N|k|\sqrt{\frac{2\tau_D}{\tau_\phi}}},
\end{equation}
i.e. at hight temperatures $\log|\delta G|$ scales with $N$ as
$N^{3/2}$ while at low temperatures it scales as $N$. The
temperature dependence of the first three harmonics of AB
conductance in the presence of electron-electron interactions is
depicted in Fig. 12.

\begin{figure}[t]
 \centering
\includegraphics[width=2.5in]{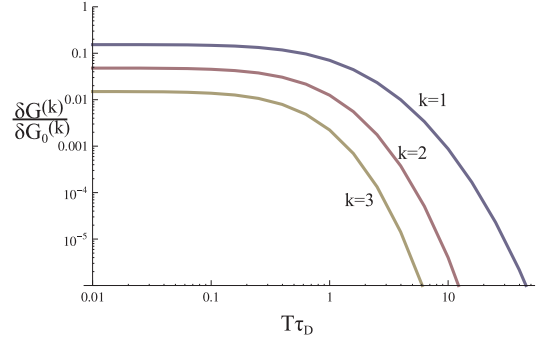}
\caption{\label{f6}Temperature dependence of the first three
harmonics of AB conductance for $g_t=500$, $g=30$, $N=10$,
$\beta_t=1$ and $\tau_D/\tau_{RC}=120$.}
\end{figure}

The results obtained here allow to formulate quantitative
predictions regading the effect of electron-electron interactions
on Aharonov-Bohm oscillations of conductance for a wide class of
disordered nanorings embraced by our model. Of particular interest
is the situation of large number of dots $N \gg 1$ which
essentially mimics the behavior of diffusive nanostructures. In
order to establish a direct relation to this important case it is
instructive to introduce the diffusion coefficient
$D=d^2/(2\tau_D)$ and define the electron density of states
$\nu=1/(d^3\delta)$, where $d$ is a linear dot size.  Then we
obtain with exponential accuracy:
\begin{equation}
\delta G^{(k)}\sim\left\{ \begin{array}{lc}
e^{-|k| (\mathcal L/\mathcal L_\phi)} & \qquad\qquad T\ll D/(\mathcal L d),\\
e^{-|k| (\mathcal L/\mathcal L_\phi)^{3/2}} & \qquad\qquad T\gg
D/(\mathcal Ld).
\end{array}
\right.
\nonumber
\end{equation}
Here we introduced the ring perimeter $\mathcal L=Nd$ and the
effective decoherence length
\begin{equation}
\mathcal L_\phi=\left\{ \begin{array}{lc}
 \left(\frac{\pi\nu d^3 D}{\ln\frac{4E_C}{\delta}}\right)^{1/2}& \qquad\qquad T\ll D/(\mathcal L d),\\
\left(\frac{12\nu d^2 D^2}{T}\right)^{1/3} & \qquad\qquad T\gg
D/(\mathcal L d).
\end{array}
\right. \nonumber
\end{equation}
Note in the high temperature limit $T\gg D/(\mathcal L d)$ the
above results match with those derived earlier for metallic
nanorings with the aid of different approaches \cite{TM,LM}. On
the other hand, at lower $T$ our results are different. This
difference is due to low temperature saturation of $\tau_\phi$
which was not accounted for in \cite{TM,LM}. A non-trivial
feature predicted here is that -- in contrast to weak localization
\cite{GZ} -- the crossover from thermal to quantum dephasing is
controlled by the ring perimeter $\mathcal L$. This is because
only sufficiently long electron paths fully encircling the ring
are sensitive to the magnetic flux and may contribute to AB
oscillations of conductance.

We believe that the quantum dot rings considered here can be
directly used for further experimental investigations of quantum
coherence of interacting electrons in nanoscale conductors at low
temperatures.

\section{Comparison with experiments and concluding remarks}

Turning to the experimental situation in the field, it is important to emphasize again that low temperature saturation of the electron decoherence time has been repeatedly observed in numerous experiments and is presently considered as firmly established and indisputably existing phenomenon.  Although in some cases this phenomenon can be attributed to various extrinsic mechanisms, like magnetic impurities, overheating etc., in the vast majority of cases none of such extrinsic mechanisms can reasonably account for experimental observations. On the other hand, it was demonstrated above that electron-electron interactions universally provide non-vanishing electron dephasing down to
$T=0$ in all types of disordered conductors. Therefore, it would be interesting to perform quantitative comparison between our universal formula for $\tau_{\varphi 0}$, Eq. (\ref{t1}), and experimental values
of the electron decoherence time measured in different structures.

Note that in some of our earlier publications
\cite{GZ,Lammi,GZ02} we have already demonstrated a good
quantitative agreement between our theoretical predictions
\cite{GZ} and experimental data for $\tau_{\varphi 0}$
obtained for numerous metallic wires and quasi-1d semiconductors. Here
we address the experiments on quantum dot structures as well as on
both weakly and highly disordered metals.

First turning to quantum dots, we recall that in all 14 samples reported in experiments with open quantum dots performed by different groups \cite{Bird1,Clarke,pivin,Huibers,Hackens} the values
$\tau_{\varphi 0}$ were found to rather closely follow a simple
dependence \cite{Hackens}
\begin{equation}
\tau_{\varphi 0}\approx\tau_D.
\label{scdots}
\end{equation}
This approximate scaling was observed within the interval of
dwell times $\tau_D$ of about 3 decades, see Fig. 5 in
\cite{Hackens}. Our Eq. (\ref{t1}) essentially reproduces
this scaling, especially having in mind that the dimensionless conductance $g$ was of order one (or slighlty larger) in almost all
samples \cite{Bird1,Clarke,pivin,Huibers,Hackens}. To the best of our knowledge no alternative explanation for the scaling (\ref{scdots}) has
been offered until now. Thus, we conclude that our theory is clearly
consistent with the available experimental data on zero temperature electron dephasing in open quantum dots.

Let us now consider spatially extended disordered conductors.
As our theory of dephasing by electron-electron interactions
predicts a rather steep increase of
$\tau_{\varphi 0}$ with the system diffusion coefficient $D$, for
most weakly disordered metals as  $\tau_{\varphi 0} \propto D^3$,
we can conclude that for a large number of disordered conductors
$\tau_{\varphi 0}$ strongly {\it increases} with increasing $D$.
This trend is indeed quite obvious for relatively weakly disordered conductors.
On the other hand, Lin and coworkers \cite{BL,Lin07,Lin01,Lin07b}
analyzed numerous experimental data for $\tau_{\varphi 0}$ obtained by various groups in rather strongly disordered conductors with $D \lesssim
10$ cm$^2$/s and observed systematic {\it decrease} of
$\tau_{\varphi 0}$ with increasing $D$. The data could be fitted
by the dependence $\tau_{\varphi 0} \propto D^{-\alpha}$ with the
power $\alpha \gtrsim 1$. This trend is clearly just the {\it
opposite} to one observed in less disordered conductors with $D
\gtrsim 10$ cm$^2$/s.

In Fig. 13 we collected experimental data for $\tau_{\varphi 0}$
obtained in about 130 metallic samples with similar Fermi velocities and diffusion coefficients
varying by $\sim 3$ decades, from $D \approx 0.3$ cm$^2$/s to $D
\approx 350$ cm$^2$/s. The data were taken from about 30 different
publications listed in the figure caption. We see that the the
measured values of $\tau_{\varphi 0}$ strongly depend on $D$.
Furthermore, this dependence turns out to be non-monotonous: For
relatively weakly disordered structures with $D \gtrsim 10$
cm$^2$/s $\tau_{\varphi 0}$ increases with increasing $D$, while
for strongly disordered conductors with $D \lesssim 10$ cm$^2$/s
the opposite trend takes place. In addition to the data points in
Fig. 13 we indicate the dependencies $\tau_{\varphi 0} (D)$
(\ref{D3}) and (\ref{D-3}) for two models $(a)$ and $(b)$
discussed above.

\begin{figure}
\includegraphics[width=7.5cm]{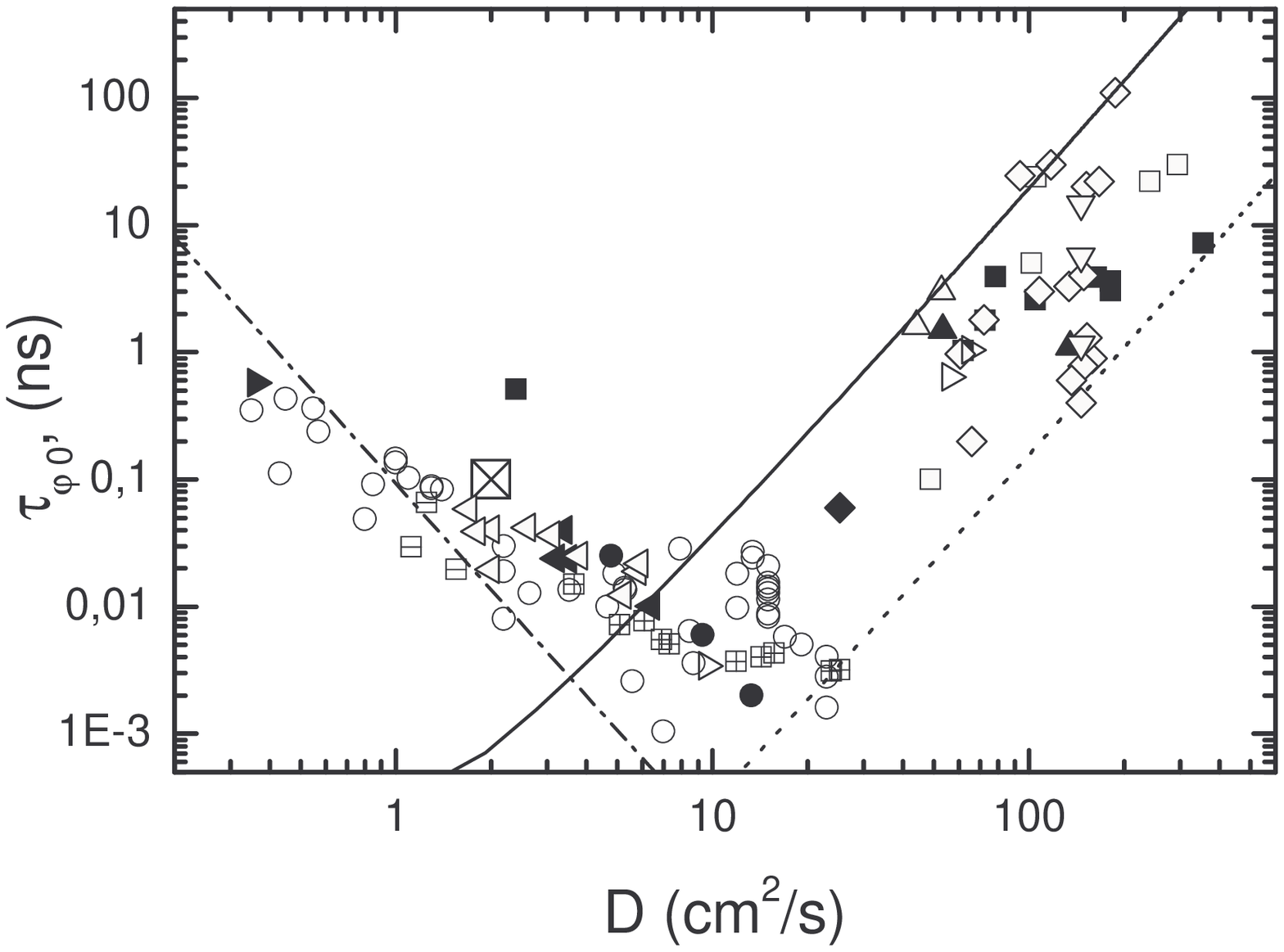}
\caption{The low temperature dephasing times observed in various
experiments for the following samples: Au-1 to Au-6 \cite{Moh},
Au-7 \cite{MJW}, Au-8 and Au-10 \cite{Mohunp} ($\blacksquare$); 44 samples (AuPd and AgPd) \cite{Lin07} ($\circ$); 18 samples \cite{Lin01}: Au$_2$Al
($\blacktriangleleft$), Sb ($\triangleleft$), Sc$_{85}$Ag$_{15}$
($\blacktriangleright$), V$_3$Al ($\boxminus$); 9 samples (CuGeAu)
\cite{Lin07b} ($\boxplus$); 15
samples (Au, Ag and Cu)  \cite{Saclay} and AgMI6N
\cite{Birgeunp} ($\lozenge$); CF-1 and CF-2 \cite{MW}
($\triangle$); A, B (Au) \cite{Ba}, Au1 \cite{Gre}, Ag1
\cite{many} and Ag2 \cite{Gre-new} ($\square$); S, M and L (Pt) \cite{Bird}
($\bullet$); D and F \cite{Nat05} ($\triangleright$); Ag, AgFe1
and AgFe2 \cite{Birge} ($\triangledown$); 10 samples \cite{Sah} within the box ($\boxtimes$); 2 (Au) \cite{Enh} and 1 (Au) \cite{Gersh93}
($\blacktriangle$);   al-1 \cite{Altomare}
($\blacklozenge$). Our Eq. (\ref{D3}) for $\gamma = 0.2$ and 1 is
indicated respectively by solid and dashed lines, while Eq.
(\ref{D-3}) for $g_c=150$ is depicted by dashed-dotted line.}
\end{figure}

We observe that for  $D \gtrsim 10$ cm$^2$/s the data points
clearly follow the scaling (\ref{D3}). Practically all data points
remain within the strip between the two lines corresponding to Eq.
(\ref{D3}) with $\gamma = 1$ (dashed line) and $\gamma = 0.2$
(solid line). On the other hand, for more disordered conductors
with $D \lesssim 10$ cm$^2$/s the data are consistent with the
scaling (\ref{D-3}) obtained within the model $(b)$. We would like
to emphasize that theoretical curves (\ref{D3}) and (\ref{D-3})
are presented in Fig. 13 with {\it no fit parameters} except for a
geometry factor $\gamma$ for the first dependence and the value
$g_c \approx 150$ for the second one. This value of $g_c$ was
estimated from the crossover condition (\ref{estD}) with $D \sim
10$ cm$^2$/s and $\gamma \sim 1$.

Now let us  consider the data for strongly
disordered conductors with  $D< 10 $ cm$^2$/s. As we already
pointed out, the agreement between the data and the dependence
(\ref{D-3}) predicted within our simple model $(b)$ is reasonable,
in particular for samples with $D< 3$ cm$^2$/s. At higher
diffusion coefficients most of the data points indicate a weaker
dependence of $\tau_{\varphi 0}$ on $D$ which appears natural in
the vicinity of the crossover to the dependence (\ref{D3}). The
best fit for the whole range 0.3 cm$^2$/s $<D< 10$ cm$^2$/s is
achieved with the function $\tau_{\varphi 0} \propto D^{-\alpha}$
with the power $\alpha \approx 1.5 \div 2$.

Thus, we conclude that our theory
allows to qualitatively understand and explain seemingly
contradicting dependencies of $\tau_{\varphi 0}$ on $D$ observed
in weakly and strongly disordered conductors. While the trend
``less disorder -- less decoherence'' (\ref{D3}) for sufficiently
clean conductors is quite obvious, the opposite trend  ``more
disorder -- less decoherence'' in strongly disordered structures
requires a comment. The latter dependence may indicate that with
increasing disorder electrons spend more time in the areas with
fluctuating in time but spatially uniform potentials. As we
already discussed in the beginning of Sec. 3, such
fluctuating potentials do not
dephase and thus $\tau_{\varphi 0}$ gets effectively increased. In other
words, in this case the corresponding dwell time $\tau_D$ in Eq.
(\ref{t1}) becomes longer with increasing disorder and, hence, the
electron decoherence time $\tau_{\varphi 0}$ does so too.

Note that since local
conductance fluctuations increase with increasing disorder,
several grains can form a cluster with internal inter-grain
conductances strongly exceeding those at its edges. In this case
fluctuating potentials remain almost uniform inside the whole
cluster which will then play a role of an effective (bigger)
grain/dot. Accordingly, the average volume of such ``composite
dots'' $\mathcal{V} \propto 1/\delta$ may grow with increasing
disorder, electrons will spend more time in these bigger dots and,
hence, the electron decoherence time (\ref{t1}) will increase.

The above comparison with experiments confirms that our previous
quasiclassical results \cite{GZ} for $\tau_{\varphi 0}$
are applicable to relatively weakly
disordered structures with $D\gtrsim 10$ cm$^2$/s, while for
conductors with stronger disorder different expressions for
$\tau_{\varphi 0}$ (e.g., Eq. (\ref{D-3})) should be used.
Our analysis also allows to rule out scattering on magnetic
impurities as a cause of low temperature saturation of
$\tau_\varphi$. The latter mechanism can explain neither strong
and non-trivial dependence of the electron decoherence time on $D$
nor even the
level of dephasing observed in numerous experiments. E.g., in
order to be able to attribute dephasing times as short as
$\tau_{\varphi 0} \lesssim 10^{-12}$ s to magnetic impurities one
needs to assume huge concentration of such impurities ranging from
few {\it hundreds} to few {\it thousands} ppm which appears highly
unrealistic, in particular for systems like carbon nanotubes,
2DEGs or quantum dots. Similar arguments were independently
put forward by Lin and coworkers \cite{Lin07,Lin07b}.

Thus, although electron dephasing due to scattering on magnetic
impurities is by itself an interesting issue, its role in low
temperature saturation of $\tau_\varphi$ in disordered conductors
is sometimes strongly overemphasized. Since the latter phenomenon
has been repeatedly observed in {\it all} types of disordered
conductors, the physics behind it should most likely be universal
and fundamental. We believe -- and have demonstrated here -- that
it is indeed the case: Zero temperature electron decoherence in
all types of conductors discussed above is caused by
electron-electron interactions.

This work was supported in part by RFBR grant 09-02-00886.

\end{document}